\documentclass[graybox]{svmult}

\usepackage{type1cm}        
\usepackage{makeidx}         
\usepackage{graphicx}        
\usepackage{multicol}        
\usepackage[bottom]{footmisc}
\usepackage{hyperref}
\usepackage{newtxtext}       %
\usepackage{newtxmath}       
\usepackage[capitalise]{cleveref}
\usepackage{caption, subcaption}

\newcommand{\Hy} {\mathscr H}
\newcommand{\Gr} {\mathscr G}
\newcommand{\Lap}{\mathscr L}
\newcommand{\R}{\mathbb R}

\makeindex             

\usepackage[normalem]{ulem}

\begin{document}

\title*{Consensus Dynamics and Opinion Formation on Hypergraphs}
\author{Leonie Neuhäuser, Renaud Lambiotte, Michael T. Schaub}
\institute{Leonie Neuhäuser  \at RWTH Aachen University, Germany, \email{neuhaeuser@cs.rwth-aachen.de},\newline Renaud Lambiotte \at University of Oxford, UK, \email{lambiotte@maths.ox.ac.uk},\newline Michael T. Schaub  \at RWTH Aachen University, Germany, \email{schaub@cs.rwth-aachen.de}}

%
%
\maketitle

\abstract{In this chapter, we  derive and analyse models for consensus dynamics on hypergraphs. 
As we discuss, unless there are \emph{nonlinear} node interaction functions, it is always possible to rewrite the system in terms of a new network of effective pairwise node interactions, regardless of the initially underlying multi-way interaction structure. 
We thus focus on dynamics based on a certain class of non-linear interaction functions, which can model different  sociological phenomena such as peer pressure and stubbornness. 
Unlike for linear consensus dynamics on networks, we show how our nonlinear model dynamics can cause shifts away from the average system state. 
We examine how these shifts are influenced by the distribution of the initial states, the underlying hypergraph structure and different forms of non-linear scaling of the node interaction function. 
}
\vspace{1cm}
\small The exposition in this manuscript is based on our exposition in Neuhäuser L., Mellor A., Lambiotte R.: Multibody interactions and nonlinear consensus dynamics on networked systems. Phys. Rev. E, 101 (3) (2020),
Neuhäuser L., Schaub M. T., Mellor A., Lambiotte R.: Opinion Dynamics with Multi-Body Interactions. arXiv preprint arXiv:2004.00901 (accepted to NETGCOOP 2021) and
Sahasrabuddhe R., Neuhäuser L., Lambiotte R.: Modelling Non-Linear Consensus Dynamics on Hypergraphs. Journal of Physics: Complexity (2021)
and reuses some of the text and results presented there.
Here we provide a unified presentation of the results presented in these publications.

\section{Background: Modelling Group Interactions}
\label{LiteratureGroups}
\begin{figure}
\centering
\includegraphics[width=\linewidth]{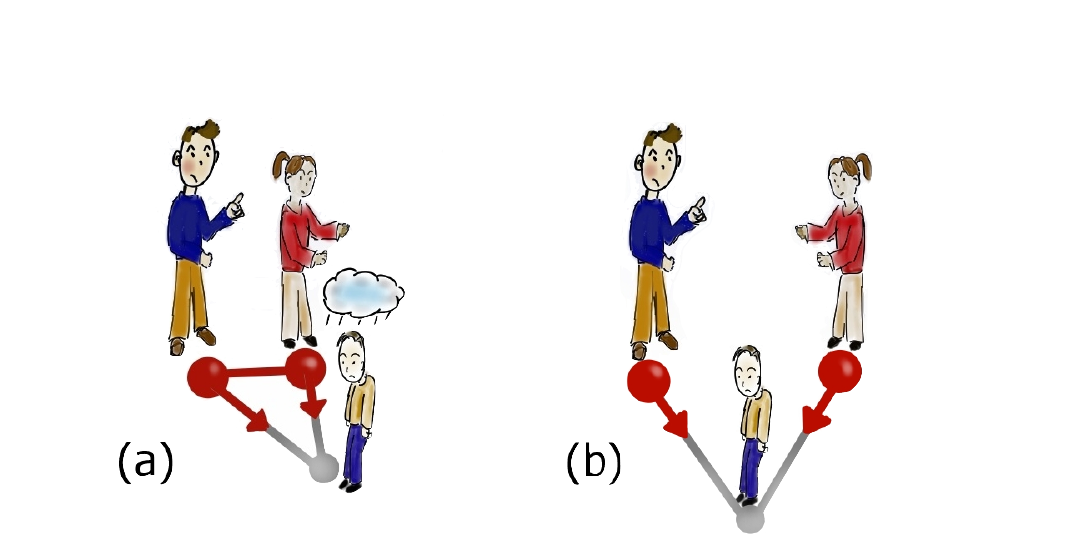}
\caption{\textbf{Higher-order group interactions in social context. } Higher-order group interactions (a) can result in greater influence on target nodes than (b) pairwise interactions.}
\label{socialcontext}
\end{figure}

Group interactions are present in various areas in nature \cite{santos_topological_2019}, society \cite{iacopini_simplicial_2019} and technology \cite{olfati-saber_consensus_2007}. Examples range from collaborations of authors \cite{patania_shape_2017} to neuronal activity \cite{giusti_clique_2015,reimann_cliques_2017}. 
In sociology, for instance, it is well known that the dynamics in a social clique is determined not just by the pairwise relationships of its members, but often by complex mechanisms of peer influence and reinforcement \cite{petri_simplicial_2018}. 
This is illustrated by the example of joint parental discipline shown in Figure \ref{socialcontext}. In Figure \ref{socialcontext}(a), a strong link between the parents reinforces the influences dynamics resulting in a stronger effect on the child. This is not captured by the independent pairwise influences in Figure \ref{socialcontext}(b).

In the context of modelling such multi-way dynamics, it is thus important to distinguish between the pairwise interactions between individuals and higher-order interactions, which cannot be decomposed further into pairwise interactions.
Specifically, if the influence on an agent can be fully explained by its pairwise relationships to other group members, then the system can be abstracted by an (effective, derived) pairwise network representation.
In contrast, higher-order interactions account for the effect of the group as a whole, and thus different frameworks than graphs are required to encode the interactions between agents. 

Especially complex social processes such as the adoption of norms or opinion spreading might not be explainable by a simple exchange of the states of neighbouring nodes, as simple models for, e.g., epidemic spreading would suggest. 
For instance, experiments in social psychology such as the conformity experiment~\cite{asch_effects_1951} indicate that multiple exposures might be necessary for an agent to adopt a certain state. 
This type of behaviour is also at the core of threshold models on networks, which model adoption processes (e.g. opinion spreading) in social systems. 
A threshold model posits that each node in a network has an associated binary state, and the (binary) state of agents only switches if a certain fraction (or a certain number) of their neighbours agrees on the same opinion  \cite{watts2002simple,lambiotte_dynamics_2008}.
More generally, such a nonlinear dependence of a node on all its neighbors may be captured via a generalized linear model, in which each node is influenced according to a nonlinear map applied to a linear transformation of the states of its neighbors, i.e., we first linearly accumulate pairwise influences and then transform the result in a nonlinear way \cite{nelder_generalized_1972}.
For instance, in the case of the threshold model this nonlinear function is a threshold or Heaviside function, which is applied to the (linear) mean of the neighbors' opinions. 
While generalised linear models can capture certain aspects of a group dynamics, these models may nonetheless provide an over-simplified view of the system. 
Consider again the previous example of parental discipline. 
If we were to model this situation with a threshold model, then any pair of adults could influence the child in the same way, irrespective of the relation between the adults.
However, it is not difficult to imagine that if there is strong relation between the parents, the influence on the child may be stronger.
We are therefore interested in models that can capture \emph{multi-way} relations that cannot be encoded with a network of pairwise influences.

There are different ways to encode such multi-way group relations in a network, including set systems, general hypergraphs \cite{noauthor_book_1987} and simplicial complexes~\cite{schaub_random_2018,mukherjee_random_2016,parzanchevski_simplicial_2017,muhammad_control_nodate,petri_simplicial_2018, masulli_algebro-topological_nodate}. 
Note that we are using the term ``network'' in the sense of system and not as a synonym for a graph.
In this chapter, we explore dynamical models based on hypergraphs, where each node has an associated state and the evolution of those states depends on the values of all the nodes inside each hyperedge.

\label{sec:introduction}
\section{From pairwise to multi-way interactions}

Pairwise dynamical systems can describe a wide range of nonlinear dynamics on graphs~\cite{vasile_sen_2015}. 
We can formalize these dynamical systems as follows.
Let $\Gr$ be a (directed) graph consisting of a set $V(\Gr)=\{1,\dots,N\}$ of $N$ nodes connected by a set of edges $E(\Gr)=\{(i,j): i,j \in V(\Gr)\}$, described by ordered tuples of nodes. 
The structure of the network can  be  represented by the adjacency matrix $A \in  \mathbb{R}^{N \times N}$ with entries
\begin{align}
	A_{ij}=\begin{cases}1 & (i,j) \in E(\Gr ) \\
		0 & \text{ otherwise}.\end{cases}
\end{align}
For simplicity we will consider only undirected networks in this chapter, in which case the adjacency matrix $A$ is symmetric.

We endow each node $i \in  V (\Gr)$  with a dynamical variable, $x_i \in \mathbb{R}$.
For a pairwise dynamical system as we consider here, the evolution of these variables is mediated by the underlying graph $\Gr$, whose edges constrain which nodes can interact with each other, and a set of node interaction functions 
\begin{align}
\label{eqn:twoway}
	\mathcal{F}=\left\{f_{i j} \; \big\vert \; f_{i j} : \mathbb{R}^2 \rightarrow \mathbb{R},(i, j) \in E(\Gr) \right \},
\end{align}
that quantify how the states of neighbouring nodes affect each other. 
Given the information about $\Gr$ and $\mathcal{F}$, the time-evolution of the pairwise system is then defined as
\begin{align}
\label{eqn:pairwise_interaction}
	\dot{x}_{i}=\sum_{j =1}^N A_{ij}f_{i j}\left(x_{i}, x_{j}\right),
\end{align}
where each node is affected by the sum of possibly nonlinear interactions with its neighbours. 
Particular examples of (\ref{eqn:pairwise_interaction}) include the Kuramoto model~\cite{arenas2008synchronization}, continuous-time random walks and linear consensus on networks \cite{masuda_random_2017}.\footnote{Note that the case of generalized linear models mentioned in the introduction does \emph{not} fit the above dsecription. Instead of consisting  of a linear sum of possibly nonlinear functions, a generalized linear model would be of the form $\dot{x}_i = f(\sum_j A_{ij} x_j)$, a nonlinear function applied to a (linear) sum.}

\subsection*{Multi-way interactions on hypergraphs}		

To encode possible multi-way interactions in a dynamical system we use a hypergraph $\Hy$.
A hypergraph $\Hy$ consist of a set $V(\Hy) = \{ 1, 2, \hdots, N \}$ of $N$ nodes, and a set $E(\Hy) = \{ E_1, E_2, \hdots, E_M \}$ of $M$ hyperedges. 
Each hyperedge $E_\alpha$ is a subset of the nodes, i.e. $E_\alpha \subseteq V(\Hy)$ for all $\alpha =1,2,\hdots,M$, where each hyperedge may have a different cardinality $n_\alpha=|E_\alpha|$. 
A graph is thus simply a hypergraph constrained to contain only $2$-edges.
We use $E^{n}(\Hy)$ to denote the set of all hyperedges of cardinality $n$, which are henceforth referred to as $n$-edges.
In the following we will concentrate on hypegraphs without self-loops, i.e., $n_\alpha \ge 2$ for all hyperedges $E_\alpha$.

We can describe the structure of a hypergraph $\Hy$ by a set of adjacency tensors ${\{A^{(n)}, n=2,3\hdots,N\}}$, where each tensor $A^{(n)}$ represents the connections made by $n$-edges.
\begin{equation}
    A^{(n)}_{ij\hdots} = \begin{cases}
1 & \{ i, j\hdots \} \in E^n(\Hy)\\
0 & \text{otherwise}
\end{cases}
\end{equation}	
Thus the adjacency tensor $A^{(n)}$ is symmetric with respect to any permutation of its indices, and as we do not allow for self-loops its entries $A^{(n)}_{ij\hdots}$ can only be nonzero if all indices are distinct.

Generalising Eq.~\eqref{eqn:pairwise_interaction}, a multi-way dynamical system is now defined by a hypergraph $\mathcal{H}$ encoding the structure of the interactions between the nodes, and by a set of node interaction functions
\begin{align}
\label{eqn:multiway}
	F=\left\{f^{E_\alpha} \;\big\vert\; f^{ E_\alpha} : \mathbb{R}^{N\times N\times \ldots\times N} \rightarrow \mathbb{R}, \; E_\alpha \in E(\Hy) \right \}.
\end{align}
Analogous to the definition of $E^{n}(\Hy)$, we can also partition the function set $F$ into subsets $F^n$ based on the cardinality of the hyperedges, which then can be matched to the corresponding adjacency tensors~$A^{(n)}$:
\begin{align}
	F^n=\left\{f^{\{  i,j\hdots \} } \;\big \vert\; \{ i, j\hdots \}  \in E^n(\Hy) \right \}.
    \label{multiway_definition}
\end{align}
The set of all node interaction functions is then given by the union $F = \bigcup_{n=2}^N F^n$. 
Finally, we assume that the time-evolution of system is governed via a linear combination of the effects of each hyperedge in which a node is involved (possibly in a nonlinear way):
	\begin{equation}
        \dot{x}_i =   \sum_{n=2}^N \sum_{j,k,\hdots} A_{ijk\hdots}^{(n)}  f^{\{ i,j\hdots \} }(x_i,x_j, \hdots)
	\end{equation}

\label{sec:model}
\section{Higher-order effects and nonlinearity}
\label{sec:interaction_function}

In this section our main goal is to examine under what conditions a multi-way dynamical system as described by \eqref{eqn:multiway} cannot be rewritten as an appropriately defined pairwise dynamical system \eqref{eqn:twoway}, that is when a multi-body formalism is truly necessary to capture the complexity of a system. We tackle this problem for classes of node interaction functions satisfying desirable symmetries, described as follows.

\subsection{Symmetries and quasilinearity}
As is often the case for models of non-linear consensus or synchronisation on standard networks, we would like our model to be invariant to translation and rotation.
This is a reasonable assumption for physical and sociological interaction processes ensuring independence on the global reference frame.
A function is rotational and translational invariant if it is invariant under application of elements from the special Euclidean group $\text{SE}(N)$, which is defined as the symmetry group of all translations and rotations around the origin.  As we restrict the scope to scalar values $x_i$ on nodes, and do not consider vectors here, the rotational invariance simply means invariance under a change of signs of the values.
In the case of two-body dynamical systems, it is known that a necessary and sufficient condition for these symmetries to be satisfied is the quasi-linearity of the interaction \cite{vasile_sen_2015}, that is 
\begin{align}
\label{eqn:quasi-linearity}
\dot{x}_i =\sum_{j}A_{ij}  \, k_{ij}(|x_j-x_i|)  \, (x_j-x_i),
\end{align}
where $k_{ij}$ is an arbitrary function from $\mathbb{R}$ to $\mathbb{R}$. This form implies that the node interaction function is, for each edge, an odd function of $(x_j-x_i)$, which is a popular choice in the study of non-linear consensus \cite{Srivastava2011}. Within the language of non-linear consensus, this model belongs to the family of relative non-linear flow. 
While we cannot generally transfer these results to multi-body dynamical systems, they will provide us a guide on how to define a `minimal non-linear' model in that case.

\subsection{Linear Dynamics and Motif Matrices}
\label{sec:linear_dynamics}
We first investigate the relations between pairwise and multi-way dynamical systems in the case of linear node interaction functions.
Linear dynamics are crucial for modeling a range of different phenomena and serve as a first approximation for many nonlinear systems. 
With pairwise dynamical systems, the node interaction function is given by $f_{ij}(x_i,x_j)=c(x_j-x_i)$ where $c\in\mathbb{R}$ is a scaling constant, and the 
resulting dynamics reads
\begin{align}
\label{eq:lap}
	\dot{x}_i=\sum_{j}A_{ij}c(x_j-x_i)=- c \sum_{j} L_{ij}x_j,
\end{align}
 \label{eqn:linear_two-way}
where $L_{ij}=D_{ij}-A_{ij}$ is the network Laplacian.
Here the degree matrix $D_{ij}=\delta_{ij}d_i$ is a diagonal matrix of the degrees $d_i=\sum_{j}A_{ij}$. 
Eq. (\ref{eq:lap}) naturally arises when modelling continuous-time random walks on networks \cite{masuda_random_2017}, but also appears in the context of opinion-formation and decentralized consensus, as in the continuous-time DeGroot model \cite{olfati2007consensus}. 
For undirected, connected networks, the dynamics asymptotically converges to an average consensus $\lim_{t\rightarrow\infty} x(t) = \mathbf{1} \alpha$ for some $\alpha\in \mathbb{R}$, with a convergence rate determined by the second dominant eigenvalue of the Laplacian.

With multi-way interaction systems, the linear node interaction function is given by
\begin{equation}
	\begin{split}\dot{x}_i &= \sum_{n=2}^N \frac{1}{(n-1)!} \sum_{jk\hdots} A_{ijk\hdots}^{(n)} c\left( x_j - x_i + x_k - x_i + \hdots \right) \\
	&= \sum_{n=2}^N \frac{c}{(n-2)!} \sum_{jk\hdots} A_{ijk\hdots}^{(n)} \left(x_j - x_i\right)  \\
    &=- c \sum_{n=2}^N\sum_j L_{ij}^{(n)} x_j.
    \end{split}
\end{equation}
\label{eqn:linear_multi-way}
Here, we generalise \eqref{eqn:linear_two-way}, scaled according to the symmetry of the linear multi-way interactions on the n-edges. We have defined the \emph{motif Laplacian} for fully connected $n$-cliques as:
\begin{align}
	L^{(n)}=D^{(n)} - W^{(n)}
\end{align}
which is simply the standard Laplacian for a graph with adjacency matrix 
\begin{align}
\label{eqn:rescale}
W_{ij}^{(n)} &=   \frac{1}{(n-2)!} \sum_{kl\hdots} A_{ijk\hdots}^{(n)}.
\end{align}
This rescaled network is thus obtained by weighting  each edge by the number of n-edges to which it belongs\footnote{This is a standard procedure to project a hypergraph on a weighted network, but other possibilities exist, for instance based on the dynamics of biased random walkers \cite{carletti2020random}}. 
Eq. (\ref{eqn:linear_multi-way}) can now be written as:
	\begin{equation}
	\dot{x}_i =  - c\sum_j \Lap_{ij} x_j 
	\end{equation}
where the Laplacians for all hyperedge cardinalities are summed up to one Laplacian $\Lap_{ij} = \sum_{n=2}^N L_{ij}^{(n)}$. 
    
In other words, a multi-way dynamical system can be rewritten as a pairwise dynamical system in the case of linear dynamics, after a proper rescaling of the adjacency matrix.
This observation reveals that a genuine multi-way dynamics on hypergraphs requires a non-linear node interaction function.
Hence, linear multi-way interactions are not sufficient to produce dynamics that cannot be reduced to pairwise dynamical systems. It is therefore essential to not simply study the hypergraph structure in isolation, but also consider the dynamical process evolving on top of this hypergraph.\\[0.1cm]

\noindent\textbf{Remark} Before exploring the dynamics of multi-way systems further, let us remark upon a connection between the operation \eqref{eqn:rescale} and the `motif matrix' used to uncover communities in higher-order networks \cite{benson_higher_order_2016}. 
A \textit{motif} on $k$ nodes is defined by a tuple $(B, P)$, where $B$ is a $k\times k$ binary matrix that encodes the edge pattern between the $k$ nodes, and $P \subset \{1, 2, .\dots , k\}$ is a set of anchor nodes.
The study of such motifs is an important objective in network science~\cite{milo_network_2002}.
In Ref. \cite{benson_higher_order_2016}, the authors define a generalisation of conductance and cut, based on motifs rather than edges.
In this context they define the \emph{motif adjacency matrix} 
\begin{align}
	(W_M)_{ij} = \text{number of instances of motifs in }M  \text{ containing } i \text{ and }j, \nonumber 
\end{align}
from which a motif Laplacian could be defined. 
Equation \eqref{eqn:rescale} provides a dynamical interpretation of this quantity for $n$-edges.
An interesting path for future research could be to employ such motif Laplacians to extend random-walk based community detection techniques such as the Map equation \cite{rosvall2009map} and Markov stability \cite{delvenne2010stability,lambiotte2014random,Schaub2012} to higher-order networks.

\label{sec:Linear}
\section{Non-linear consensus dynamics on hypergraphs}
\label{sec:higher_order_effects}

In this section we explore how a non-linear node interaction function $f$ can lead to higher-order effects that do not exist in a pairwise setting. 
In particular, we investigate non-linear consensus dynamics on hypergraphs.

\subsection{Non-linear Consensus Dynamics in three-way interactions (3CM)}

As previously mentioned, group effects that cannot be reduced to pairwise interactions appear in various contexts.
In the area of sociology, reinforcing group effects such as peer pressure are a long-standing area of study, for instance in social psychology \cite{asch_effects_1951}.
It is thus important to develop models that capture these multi-way mechanisms to better understand phenomena such as hate communities \cite{johnson_hidden_2019}, echo chambers and polarisation \cite{bail_exposure_2018} in society.

Motivated by these observations, and inspired by Eq.~\eqref{eqn:pairwise_interaction}, we first consider the case when all hyperedges have the same size 3, and we introduce a three-way consensus model (3CM)\cite{neuhauser_multibody_2020} with a non-linear node interaction function of a specific form, which is akin to a consensus dynamics with group reinforcement:
\begin{align}
	f_i^{\{j,k\}}(x_i,x_j,x_k) =  \,s \left( \left| x_j-x_k\right|\right) \,\left((x_j-x_i)+(x_k-x_i)\right).
    \label{eqn:our_function}
\end{align}
where we assume the function on each $3$-edge is the same, for the sake of simplicity\footnote{Note that we adapted the notation of a multi-way node interaction as given in Eq. \eqref{multiway_definition} to emphasise that this function is symmetric in $j$ and $k$ and influencing node $i$.}.
This expression models, for each 3-edge $\{i,j,k\}$, the multi-way influence of nodes $j$ and $k$ on node $i$ by the standard linear term $\left((x_j-x_i)+(x_k-x_i)\right)$ 
modulated by a scaling function $s\left(\left|x_j-x_k\right|\right)$ of their state differences.
If the scaling function $s(x)$ is monotonically decreasing,  the influence of $j$ and $k$ on $i$  is increased if $j$ and $k$ have similar states. In such a situation we will say that $j$ and $k$ \emph{reinforce} each other's influence. In contrast, the joint influence is diminished if $j$ and $k$ have very different states which will be called \emph{inhibiting}.
This property is reminiscent of non-linear voter models in the case of discrete dynamics \cite{lambiotte_dynamics_2008,molofsky_local_1999, mellor2017heterogeneous}, where the voters change opinion with a probability $p$ that depends non-linearly on the fraction of disagreeing neighbours. 
Note that other choices of non-linear node interaction functions $f$, akin to Watts threshold model~\cite{watts2002simple}, have also been considered recently for information spreading \cite{arruda2019social}.

For the dynamics~\eqref{eqn:our_function} the resulting dynamics for each node $i$ are then given by
\begin{align}
	\dot{x}_{i}=\sum_{j , k=1}^N \frac{1}{2}A^{(3)}_{ijk} \,s \left( \left| x_j-x_k\right|\right) \,\left((x_j-x_i)+(x_k-x_i)\right).
    \label{eqn:3CM_dynamics}
\end{align}

In Eq. (\ref{eqn:our_function}), the node interaction function is non-linear for non-constant scaling functions $s(x)$ and captures multi-way effects, as the interactions on a triangle can no longer be split into pairwise node interaction functions.
If the scaling function $s(x)$ is constant, we recover the linear dynamics discussed in \Cref{sec:linear_dynamics}.

The functional form of our model has some further symmetries.
In particular, we remark that~\eqref{eqn:our_function} is invariant to translation ($x_i \rightarrow x_i+a$ for $a \in \mathbb{R}$) and equivariant to reflections through the origin ($x_i \rightarrow -x_i$) of all node states. 
This is a desirable property for many opinion formation process, as it ensures that the opinion formation is only influenced by the relative position of the node states $x_i$ and independent of a specific global reference frame. 
This is still true for the more general case of vector valued states: any rotation of the node states is norm preserving, and thus $s(\|x_j-x_k\|)$  is rotational and translational invariant. 
Since the term $\left((x_j-x_i)+(x_k-x_i)\right)$ is translation invariant and linear, any translation and rotation applied to all states will leave~\eqref{eqn:our_function} equivariant, both in the case of scalar and vector valued states. 
Note that this `quasi-linearity' of the three-way node interaction function is in close correspondence to the necessary and sufficient conditions for translation and rotational invariance for pairwise interaction systems~\cite{vasile_sen_2015}.
In the following, we restrict our scope to scalar states $x_i$. 

\begin{figure}
	\centering
	\includegraphics[width=0.3\columnwidth]{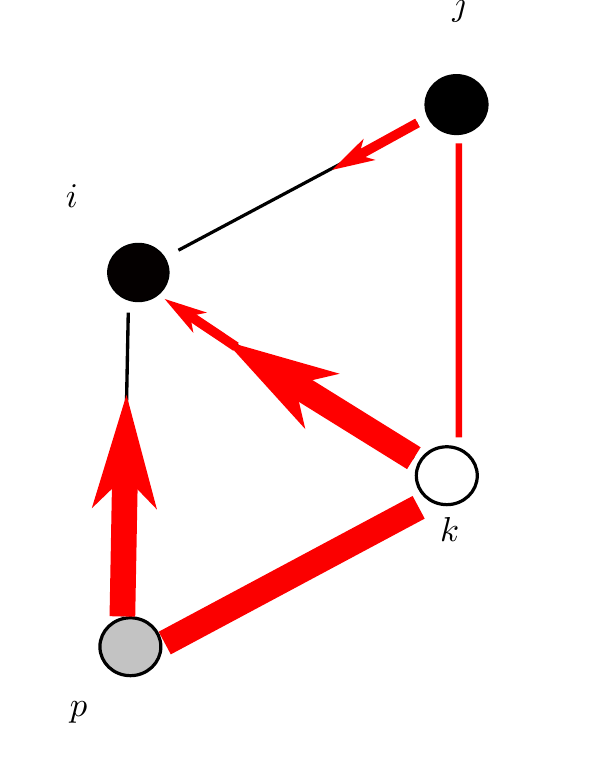}
	\caption{\textbf{Reinforcing group dynamics on 3-edges (3CM). }The influence on node $i$ due to the interactions on 3-edge $\{i,j,k\}$ and $\{i,k,p\}$. The value $x_i \in [0,1]$ of the nodes is visualised by a colour gradient between white and black. We consider a monotonically decreasing scaling function (e.g. $s(x)=\exp(\lambda x)$ for $\lambda<0$). As a result, nodes $p$ and $k$ reinforce each other as they have similar values, and hence a large $s(\left|x_p-x_k\right|)$. Nodes $k$ and $j$ have instead  distinct values, which leads to a smaller scaling function $s(\left|x_j-x_k\right|)$. Reproduced and adapted from \cite{neuhauser_multibody_2020}.
    }
	\label{fig:influence_strongandweak}
\end{figure}

As we want to be able to model a reinforcement effect for nodes with a similar opinion, a natural choice for the scaling function $s(x)$ is
\begin{align}
	\label{eqn:exponential}
	s \left( \left| x_j-x_k\right|\right)=\exp\left(\lambda \left|x_j-x_k\right|\right),
\end{align}
where the sign of the parameter $\lambda$ determines if the function monotonically decreases or increases.
Specifically, if $\lambda < 0$, then similar node states $x_j$ and $x_k$ will lead to a stronger influence on node $i$.
If $\lambda > 0$, then dissimilar node states $x_j$ and $x_k$ will lead to a stronger influence on node $i$.
Finally, if $\lambda=0$, then we recover the linear dynamics discussed above (since the scaling function $s(x)=1$ is constant).
\Cref{fig:influence_strongandweak} shows the influences on node $i$ for $\lambda < 0$, i.e. where similar node states reinforce each other.

\subsection{Derivation of a Weighted, Time-dependent Laplacian}
\label{sec:higher_order_effects:weighted_laplacian}

In \Cref{sec:linear_dynamics}, we showed that, in the case of linear node interactions, a multiway, and thus a three-way dynamical system can be rewritten as a pairwise dynamical system, defined on a network where the weight of an edge is the number of 3-nodes to which the edge belongs. 
Let us explore how this result extends to 3CM.
Recall that we assumed that the adjacency tensor $A^{(3)}_{ijk}$ is symmetric.
We define $\mathcal{I}_{ij}$ as the index-set containing all nodes $k$ that are part of $3$-edge containing node $i$ and $j$.
Note that $\mathcal{I}_{ij}=\emptyset$ if the nodes $i,j$ are not part of any 3-edge.
We now define the weighted adjacency matrix $\mathfrak{W}$ as
\begin{align}
\label{eq:hello}
(\mathfrak{W}^{(3)})_{ij}= \sum_{k}A^{(3)}_{ijk}s \left( \left| x_j-x_k\right|\right)=\sum_{k \in \mathcal{I}_{ij}}s \left( \left| x_j-x_k\right|\right).
\end{align}
Since $\mathcal{I}_{ii}=\emptyset$, because a node cannot appear more than once in a hyperedge, the weighted adjacency matrix has zero diagonal ($\text{diag}(\mathfrak{W}^{(3)})=0$).
The corresponding degree matrix, that measures the total three-way influence on node $i$  is defined as
\begin{align}
	(\mathcal{D}^{(3)})_{ii}=  \sum_{jk}A_{ijk}^{(3)}s \left( \left| x_j-x_k\right|\right) = \sum_{j} \mathfrak{W}^{(3)}_{ij}
\end{align}
and the corresponding  Laplacian is given by
\begin{align}
	\label{eqn:weighted_laplacian}
	\mathcal{L}^{(3)}= \mathcal{D}^{(3)}-\mathfrak{W}^{(3)}.
\end{align}
Using \eqref{eqn:weighted_laplacian}, we can rewrite the dynamics in \eqref{eqn:3CM_dynamics} as
\begin{align}
\label{eq:ren}
	\dot{x}_i & = \sum_{jk}\frac{1}{2}A^{(3)}_{ijk} \,s \left( \left| x_j-x_k\right|\right)\,((x_j-x_i)+(x_k-x_i)) \nonumber \\
	          & = \sum_{jk}A^{(3)}_{ijk}\, s \left( \left| x_j-x_k\right|\right)\, (x_j -x_i)           \nonumber \\
	          & = \sum_{j}\mathfrak{W}^{(3)}_{ij}(x_j-x_i)						
	=  -\sum_{j}\mathcal{L}^{(3)}_{ij}x_j.											
\end{align}
The 3CM can thus also be rewritten in terms of the Laplacian of a network represented by $\mathfrak{W}^{(3)}$, whereas the entries $(\mathfrak{W}^{(3)})_{ij}$ measure the three-way influence on node $i$ over edge $\{i,j\}$. 
However, as the adjacency matrix $\mathfrak{W}^{(3)}=\mathfrak{W}^{(3)}(t)$ depends on the dynamical node states $x_i=x_i(t)$ in (\ref{eq:hello}) and this projection is thus \emph{time-dependent} and \emph{node state-dependent}, which implies that we cannot write this dynamics in terms of pairwise interactions.
We drop these dependencies in our notation for simplicity, and simply use $\mathfrak{W}$ from now on.
The  weighted time-dependent Laplacian $\mathcal{L}^{(3)}$ is the matrix representation of the non-linear dynamics and therefore the analogue of the motif Laplacian, introduced in \Cref{sec:linear_dynamics} for linear dynamics.

\subsection{Diffusive processes on hypergraphs}
\label{subsec:General_diffusive_processes}
In order to generalise the 3CM for hypergraphs with arbitrary edge cardinality, we introduce a general model of diffusive processes on hypergraphs. Here, we consider $(x_j-x_i)$ as a diffusive coupling of nodes $i$ and $j$.
The influence of the nodes in a hyperedge $E_{\alpha}$ on a node $i$ is given by the node interaction function
\begin{equation}
	f_i^{E_\alpha}(x_i,x_j,x_k, \cdots) =  \begin{cases} 
    \sum_{j \in E_\alpha} s_{i}^{j}(x_i, x_j, x_k, \hdots)   \left(x_j - x_i\right)  & i \in E_{\alpha} \\
    0 & i \notin E_{\alpha} .
      \end{cases}
      \label{eqn:general_diffusion}
	\end{equation}
    The joint effect of all nodes within a hyperedge is given by the diffusive couplings of each node pair of influenced node ($i$) and acting node ($j$), modulated by the scaling function $s_{i}^{j}(E_{\alpha})$ which captures the influence of the hyperedge $E_{\alpha}$ as a whole.
    The scaling function $s_{i}^{j}(E_{\alpha})$ is invariant with respect to any permutation of the indices  $k \in E_\alpha$, where  $k \neq i$ and $k\neq j$. 
For the special case where it is symmetric in all node indices $k \neq i$, we can write $s_{i}({E_\alpha})$ . 

The overall dynamics of node $i$ is then obtained by linearly combining the effect of each hyperedge node $i$ is part of:
	\begin{equation}
	\dot{x}_i = \sum_\alpha f_i^{E_\alpha}(x_i,x_j,x_k, \cdots) .
	\end{equation}

For the special case of a scaling function $s_i$ which is independent of acting node $(j)$, we can derive analytical results for the behavior of this opinion formation process.
    We can then write the effect of all $n$-edges on $i$ as:
\begin{equation}\dot{x}_i = \sum_{n=2}^N \sum_{jk\hdots} A_{ijk\hdots}^{(n)} s_i(x_i, x_j, x_k, \hdots)\left(x_j - x_i\right) \times \frac{1}{(n-2)!}
	\end{equation}
    \label{eq:master_lap2}
	
    Analogously to the 3CM, we define weight matrices $\mathfrak{W}^{(n)}$and corresponding degree matrices $\mathfrak{D}^n$ as follows:
	\begin{equation}
    \label{n_weighted_matrix}
	\begin{split}
	\mathfrak{W}_{ij}^{(n)} &= \sum_{kl\hdots} A_{ijk\hdots}^{(n)} s_i(x_i, x_j, \hdots) \frac{1}{(n-2)!} \\
	\mathfrak{D}_{ii}^{(n)} &= \sum_j\mathfrak{W}_{ij}^{(n)} 
	\end{split}
	\end{equation}
	Here, $\mathfrak{D}_{ij}^{(n)} = 0$ for all $i \neq j$. Eq.(\ref{eq:master_lap2}) can now be written as:
	\begin{equation}
	\dot{x}_i = - \sum_{n=2}^N \sum_j \mathfrak{L}_{ij}^{(n)} x_j = - \sum_j \Lap_{ij} x_j 
	\end{equation}
	where $\Lap_{ij} = \sum_{n=2}^N \mathfrak{L}_{ij}^{(n)}$ and  $\mathfrak{L}_{ij}^{(n)} =\mathfrak{D}_{ij}^{(n)}  - \mathfrak{W}_{ij}^{(n)} $.
	
    We see again that when the scaling function $s_i$ is constant,  the dynamics reduces to a linear dynamics on an (effective) static weighted network as shown in Section \ref{sec:Linear}. 
    However, when the interactions are non-linear, the corresponding network is time-dependent and multi-way effects are created.

\subsection{Multi-way consensus model (MCM)}
There are multiple possibilities for specific forms of the nonlinear scaling functions $s_i^j(E_\alpha)$ in the general diffusion process in \eqref{eqn:general_diffusion}. They can have different sociological motivations and result in distinct mathematical properties, which we will show for two specific node interaction functions. They define two submodels, MCM I and MCM II, of a general Multi-way Consensus Model (MCM) \cite{sahasrabuddhe_modelling_2020}\footnote{Note that 3CM (MCM) was originally named three-body (multi-body) consensus model in \cite{neuhauser_multibody_2020,sahasrabuddhe_modelling_2020}. Here, we prefer the more descriptive name three-way consensus model, which emphasises more clearly that it is not the number of entities involved but the type of interaction between those entities (pairwise vs multi-way) which is different to classical models}. 
	\subsubsection{MCM I models homophily}
\label{ssec:MCM1}
    Homophily is a central concept in sociology  describing the tendency of like-minded individuals to interact \cite{mcpherson2001birds}. The topology of social interactions is often influenced heavily by homophily. Here, we consider the underlying topology fixed, but interpret homophily instead as a force modulating the effect of a hyperedge, depending on the proximity of the opinions inside it. 
   
   This is captured by the following node interaction function (MCM I)
       \begin{equation}\label{eq:mcm1}
	f_i^{E_\alpha}(x_i,x_j,x_k, \cdots) = s_i^I( g_i^{E_{\alpha}}(x)) \sum_{j \in E_\alpha} (x_j - x_i) ,
	\end{equation}
    where the argument function $g_i^{E_{\alpha}}$ of the scaling function $s_i^I$ measures the distance of the state of \textbf{node $i$} to the mean state of the hyperedge \textbf{including $i$}:
    \begin{align}
    g_i^{E_{\alpha}}(x_i,x_j,x_k, \cdots) =  \left |\frac{\sum_{k \in E_\alpha} x_k}{\mid E_\alpha \mid} - x_i \right | = |\langle x\rangle_{E_\alpha} -x_i| 
    \end{align}
  In sociological terms, the argument function $g_i^{E_{\alpha}}$ quantifies the difference between the opinion of individual $i$ and the average opinion of group $E_\alpha$ that $i$ belongs to. The influence of a group on a node is thus determined by the proximity of its average state to the state of the node, modulated by the scaling function $s_i^I$. For instance, a monotonically decreasing function $s^I_i$  represents an individual $i$ who is less influenced by groups with opinions very different from its own than by groups with similar opinions. 
     
Mathematically speaking, $g_i^{E_{\alpha}}$ is independent of acting node $j$ within the hyperedge, and $s_i^I(g_i^{E_{\alpha}}(x))$ thus modulates the competing effect of different hyperedges on the  state of an incident node. In other words, the scaling function  determines the rate at which a certain hyperedge influences the state of a node.
  
\subsubsection{MCM II models conformity}
\label{ssec:MCMII}
     Conformity is used to describe the tendency of an individual to align its beliefs to those of its peers, and is usually affected by the reinforcing nature of shared opinions (peer pressure). Modelling peer pressure has already been the motivation for the node interaction function of 3CM in \eqref{eqn:our_function} before. We generalise this for hyperedges of arbitrary size (MCM II):
     \begin{equation}\label{eq:mcm2}
	f_i^{E_\alpha}(x_i,x_j,x_k, \cdots)  = \sum_{j \in E_\alpha} s_i^{II}(g_{i \leftarrow j}^{E_{\alpha}}(x))(x_j - x_i).
	\end{equation}
      where the argument function $g_{i \leftarrow j}^{E_{\alpha}}$ of $s_i^{II}$ now measures the distance of the state of a \textbf{participating node $j$} from the mean state of the hyperedge \textbf{excluding $i$}.
    \begin{align}
      g_{i \leftarrow j}^{E_{\alpha}}(x_i,x_j,x_k, \cdots) = \left|\frac{\sum_{k \in E_\alpha, k \neq i} x_k}{|E_\alpha| - 1} - x_j \right | = |\langle x\rangle_{E_\alpha \setminus i} -x_j|
    \end{align}
     When all the hyperedges have size 3, we then have that $$g_{i \leftarrow j}^{\{i,j,k\}}(x_i,x_j,x_k)  = \left|\frac{x_j + x_k}{2} - x_j \right |.  $$
 and the node interaction function is thus given by
 \begin{equation}
	f_i^{\{i,j,k\}}(x_i,x_j,x_k)  = s_i^{II}\left(\frac{\left| x_k - x_j \right |}{2}\right) \left( \left (x_j - x_i \right ) + \left (x_k-x_i \right) \right).
	\end{equation}
    which indeed recovers the node interaction function \eqref{eqn:our_function} of 3CM, with a constant that can be absorbed in the time scale.
    
   In sociological terms, the argument function $g_{i \leftarrow j}^{E_{\alpha}}$  of $s^{II}_i$ captures the difference between the opinion of individual $j$ to the average opinion of the group except individual $i$. Thus, in MCM II the influence exerted by $j$ inside a hyperedge depends on the proximity of its opinion to those of the rest of the group. 
     
 Mathematically,  $g_{i \leftarrow j}^{E_{\alpha}}$ is dependent on acting node $(j)$ and $s^{II}_i ( g_{i \leftarrow j}^{E_{\alpha}}(x))$ thus determines which nodes inside a single hyperedge are the most influential.

  Like in 3CM, for a monotonically decreasing function an individual $i$ tends to be more influenced by individuals who agree with the rest of the group. For an increasing function,  individuals are more attracted to the outliers of a group (anti-conformists or contrarians).
  	
  It becomes clear here that even if we choose $s(x) = s_i^I(x) = s_i^{II}(x)$ to be of the same form, the behaviour of the two facets will differ due to their arguments.
 \subsubsection{General Multi-way consensus model}
We can combine both of these submodels in a more general Multi-way Consensus Model (MCM) \cite{sahasrabuddhe_modelling_2020}, which can capture both homophily and conformity. The overall effect of the hyperedge $E_\alpha$ on node $i \in E_\alpha$, is then given by
	\begin{equation}
	\begin{split}
f_i^{E_\alpha}(x_i,x_j,x_k, \cdots)  &= s_i^I( g_i^{E_{\alpha}}(x))\sum_{j \in E_\alpha}  s_i^{II}(g_{i \leftarrow j}^{E_{\alpha}}(x))(x_j - x_i) 
	\end{split}
	\end{equation}
    We note that, as in the case of 3CM, both node interaction functions are invariant under translations ($x_i \mapsto x_i + a$ for $a \in \R$) and reflection through the origin ($x_i \mapsto - x_i$). Thus, the dynamics are independent of the frame of reference in $\R$.

\label{sec:Nonlinear}

\section{Opinion drifts: Higher-order effects of non-linearity }
In the previous section we saw that non-linearity of the dynamics is important to make genuine higher-order effects appear, which can not be explained by pairwise interactions.  This emphasises that in the non-linear cases, we have to pay extra attention to how the interaction of dynamics and higher-order topology affects the overall dynamics of the system. Therefore, our primary objective in this section is to determine if, and how, non-linear consensus models on hypergraphs asymptotically reach consensus and which aspects can influence the dynamics.
We can do that by identifying and  analysing conserved quantities, which is usually an essential step to understand the properties of dynamical systems.

\subsection{Conservation and shifts of the average node state}
\label{sec:higher_order_effects:conservation}
In particular, we will investigate the average node state of the system. 
For consensus dynamics on graphs, it is well-known that the average state at time $t$, 
\begin{align}
	\bar{x}(t)=\frac{1}{N}\sum_{i=1}^Nx_i(t),
\end{align}
is conserved under general conditions. 
Specifically, consider a pairwise dynamical system described by
\begin{align}
	\dot{x}_i(t) &= \sum_{j=1}^N A_{ij}f_{ij}(x_i(t),x_j(t)) = \sum_{j=1}^N A_{ij}h(x_j(t)-x_i(t)). \nonumber
\end{align}
The initial average $\bar{x}(0)$ is conserved if the derivative $\dot{\bar{x}}(t) =\frac{1}{N}\sum_{i,j=1}^N A_{ij}h(x_j(t)-x_i(t))$ is zero for all times.
This is true if the adjacency matrix $A_{ij}$ of the graph is symmetric and the node interaction function $h(x)$ is odd, which is fulfilled by quasi-linear dynamics \eqref{eqn:quasi-linearity}.

We now investigate how these conditions for conservation of the average state change for a general multi-way interaction system. 
We consider a general node interaction function $f(x_1(t), \dots, x_n(t))=h(\sum_{j\neq i}(x_j(t)-x_i(t)))$, where the form of the node interaction function  $h$ ensures that the dynamics influence node $i$.
To determine conditions for a conservation of the average state we write
\begin{align}
\label{eqn:conservation_multi-way}
	\dot{x}_i(t) = \sum_{jk\dots}A_{ijk\dots}h\left(\sum_{j\neq i}(x_j(t)-x_i(t))\right).
\end{align}
Let $\Pi(i,j,k, \dots)$ be the set of all permutations of the $n$ indices.
Using this notation, we can conclude that the derivative 
\begin{equation}
    \dot{\bar{x}}(t)= \frac{1}{N}\sum_{i,j,k, \dots=1}^N A_{ijk\dots}h\left(\sum_{j\neq i}(x_j(t)-x_i(t))\right) 
\end{equation}
is zero for all times, if we have $A_{\pi}=A_{\tau}$ for all permutations $\pi,\tau \in \Pi(i,j,k, \dots)$ and moreover $\sum_{i=1}^{n}h(\sum_{j\neq i}x_j(t)-(n-1)x_i(t)) = 0$.
This is the case for an undirected multi-way interaction (for which $A$ is symmetric in all indices), provided $h(x)$ is a linear function.
We can thus conclude that, in line with our previous discussion, for a multi-way dynamical systems of the form \eqref{eqn:conservation_multi-way}, a linear dynamics conserves the average state of the system. For non-linear dynamics, however, we can not generally guarantee a conservation of the average state. 

\subsection{Factors influencing the consensus process for non-linear dynamics}

For concreteness let us consider a simple but illustrative case for three-way interactions to gain some intuition for the possible effects we can observe in a multi-way interaction system.
For 3CM, where the node interaction function takes the form $f(x_i,x_j,x_k)=s \left( \left| x_j-x_k\right|\right)((x_j-x_i)+(x_k-x_i))$, the change in the average state can be written as
\begin{align}
	\dot{\bar{x}}(t)= \frac{1}{N}\sum_{i=1}^N\dot{x}_i(t) & = \frac{1}{N}\sum_{i,j=1}^N \mathfrak{W}(t)_{ij}(x_j(t)-x_i(t)).
\end{align}
It is important to note that when the dynamics is non-linear,  $\mathfrak{W}(t)$ is time-dependent. Hence, the average state is only conserved if $\mathfrak{W}(t)$ is symmetric for all times. In particular, this means that for all $i,j$,
\begin{align}
	\mathfrak{W}(t)_{ij}                                           =\mathfrak{W}(t)_{ji}                                                      
	\Leftrightarrow
		\sum_{k \in \mathcal{I}_{ij}}s\left(\left|x_j(t)-x_k(t)\right|\right) & = \sum_{k \in \mathcal{I}_{ij}}s\left(\left|x_i(t)-x_k(t)\right|\right), \nonumber
	\label{eqn:symmetry}
\end{align}
where $\mathcal{I}_{ij}$ is the index-set containing all nodes $k$ that are part of $3$-edge containing node $i$ and $j$. This is only true for all times if the scaling function is constant ($s(x)=c$), i.e., when the dynamics is linear.
In this case, the weighted matrix $\mathfrak{W}_{ij}=c(W^{(3)})_{ij}$ simply equals the motif adjacency matrix scaled by the constant $c$ (cf. \Cref{sec:linear_dynamics}).

Otherwise, the weighted matrix $\mathfrak{W}(t)$ may be asymmetric at some time point $t$, which implies that the average state can shift, i.e., that we can get a drift of opinions. These possible shifts are influenced by an interplay between

\begin{enumerate}
\item the initial node states,
\item the scaling function $s(g(x))$, and
\item the hypergraph topology
\end{enumerate}

as all of these aspects are encoded in the weighted matrix $\mathfrak{W}^{(3)}$. This is also true for the more general case of non-linear diffusive dynamics on general hypergraphs, represented by the weighted matrix $\mathfrak{W}^{(n)}$\eqref{n_weighted_matrix}.

\subsection{Influence of the initial node states on the final consensus value in a fully connected hypergraph}
\label{sec:higher_order_effects:meanfield}

We first consider a system which eliminates all topological effects, given by a fully connected  hypergraph $\mathcal{H}$ with nodes $V(\mathcal{G})=\{1, \dots, N\}$, in which each $n$-tuple of distinct nodes is connected by a hyperedge. This is to investigate the isolated effect of the initial node states.

Let us first stay with the model example of the 3CM to get some intuition of the higher-order effects that appear in a fully connected hypergraph of 3-edges. As all possible hyperedges exist, the node set $\mathcal{I}_{ij}=V(\mathcal{G})/\{{i,j\}}$ is given by all nodes except $i$ and $j$ and the symmetry condition 
$$\sum_{k \in \mathcal{I}_{ij}}s\left(\left|x_j(t)-x_k(t)\right|\right)  = \sum_{k \in \mathcal{I}_{ij}}s\left(\left|x_i(t)-x_k(t)\right|\right) $$
thus implies that the multi-way effects have to balance out for all nodes in the network in order for the average state to be conserved. The equality here only depends on the initial node states and the scaling function $s(x)$, as the topology of the hypergraph captured by the node set $\mathcal{I}_{ij}$ does not have an influence in a fully connected system.

As an illustration,  consider a situation where the number of nodes is even and when the initial values $x_i(0)$ on the nodes is binary, that is either zero or one.  
The symmetry condition (\ref{eqn:symmetry}) will be satisfied only if the initial configuration $x(0)$ is balanced, that is when  $\bar{x}(0)=0.5$, in which case this average state is conserved in time. 
In contrast, if the initial configuration is unbalanced, (\ref{eqn:symmetry}) will not hold in general, and the average state will evolve in time. 
If $s \left( \left| x_i-x_j\right|\right) $ is given by a decreasing function, that is when similar node states reinforce each other, the deviation from  $0.5$ is expected to grow in time, with a drift towards the majority.
In contrast, if the scaling function is such that dissimilar node states reinforce each other, one will observe a drift to the balanced state $0.5$.

To validate these findings, we performed numerical simulations of 3CM on a fully connected 3-edge hypergraph of 100 nodes.
We used the exponential scaling function $s(x)=\exp(\lambda x)$ with the parameter $\lambda$ set to $\lambda=-1$ to obtain an example for a decreasing function, $\lambda=1$ as an example for a growing function, and $\lambda=0$ to obtain a constant function (i.e., a linear dynamics).

\begin{figure}
\centering
\begin{subfigure}{0.45\textwidth}
\includegraphics[width=\textwidth]{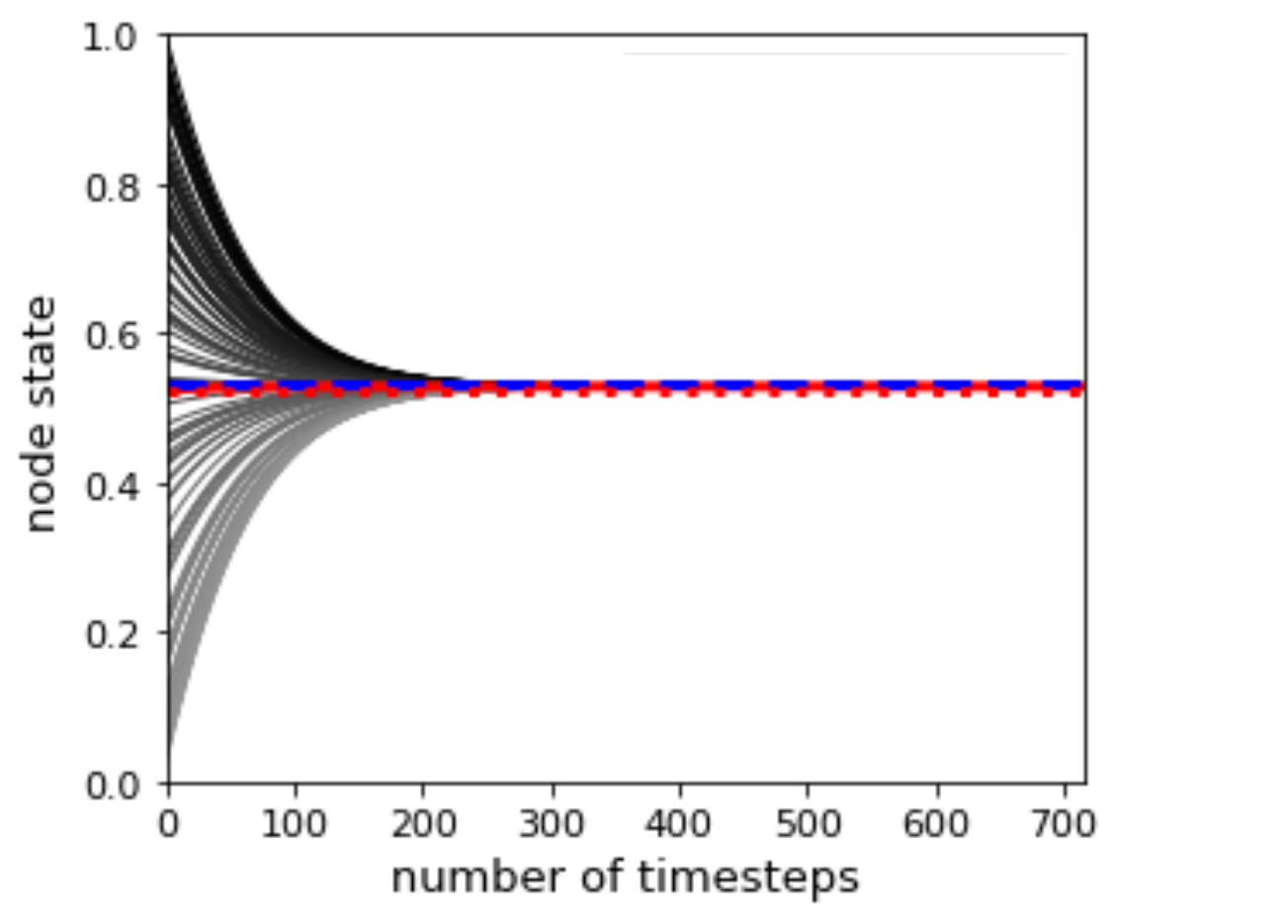}
\caption{$\mathcal{U}([0,1])$}
\end{subfigure}
\begin{subfigure}{0.45\textwidth}
\includegraphics[width=\textwidth]{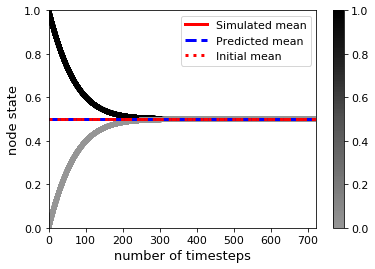}
\caption{$\mathcal{B}(0.5)$}
\end{subfigure}
\caption{\textbf{Symmetric initialisation: Conservation of average state.} For  uniform random (a) and binary (b) symmetric initialisation the average is preserved for dynamics with a scaling function $s(x)=\exp(\lambda x)$ for all $\lambda=\{1,0,-1\}$, so for both linear and non-linear dynamics. Shown are the results for $\lambda=-1$ and the line colors represent the initial states of the nodes. The predicted average agrees with the simulated average in this case. 
}
\label{fig:simulations_MeanField_symmetric}
\end{figure}

\begin{figure*}
	\centering
	\begin{subfigure}{0.32\textwidth}
		\includegraphics[width=\textwidth]{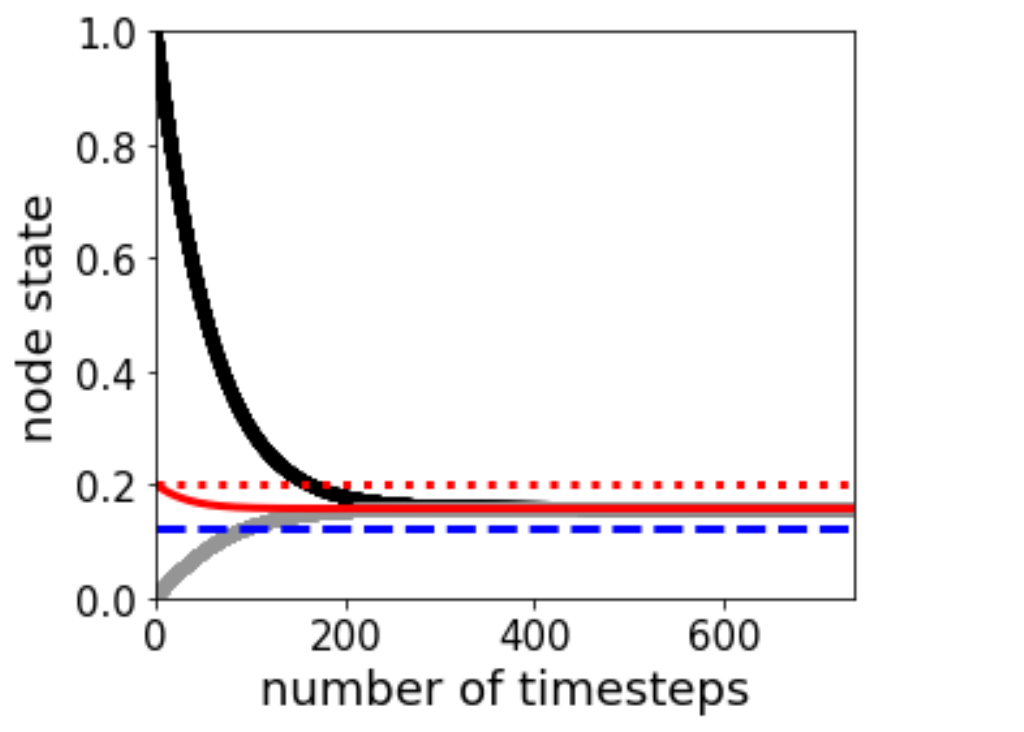}
		\caption{$\lambda=-1$ (reinforcing)}
	\end{subfigure}
	\begin{subfigure}{0.32\textwidth}
		\includegraphics[width=\textwidth]{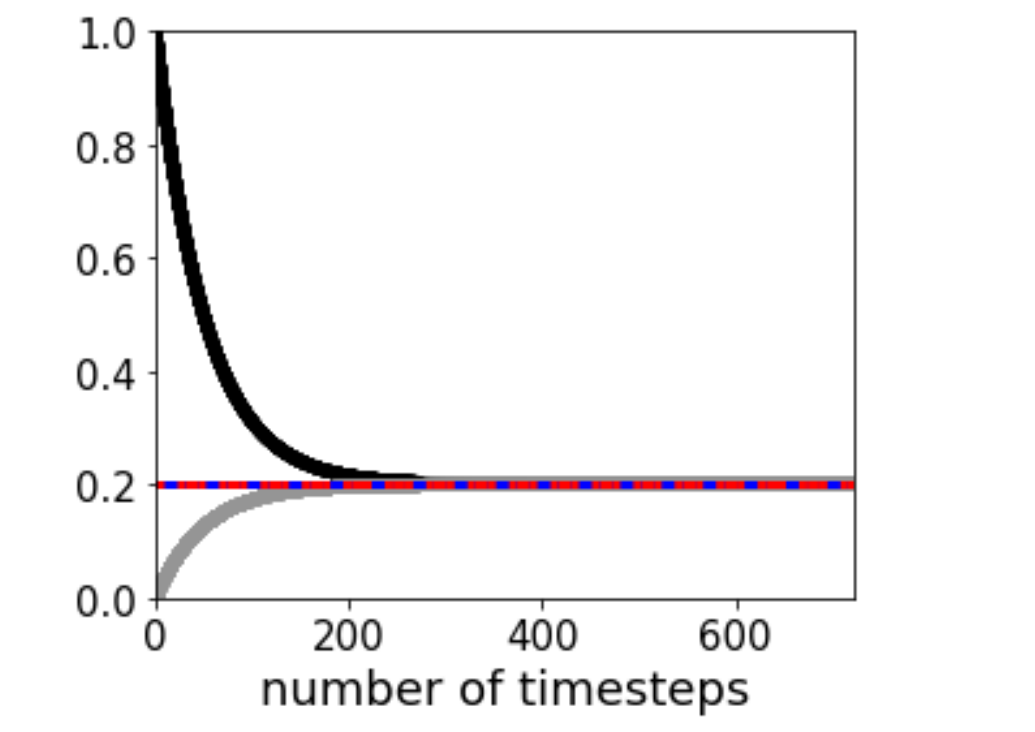}
	\caption{$\lambda=0$ (linear)}
	\end{subfigure}
	\begin{subfigure}{0.32\textwidth}
		\includegraphics[width=\textwidth]{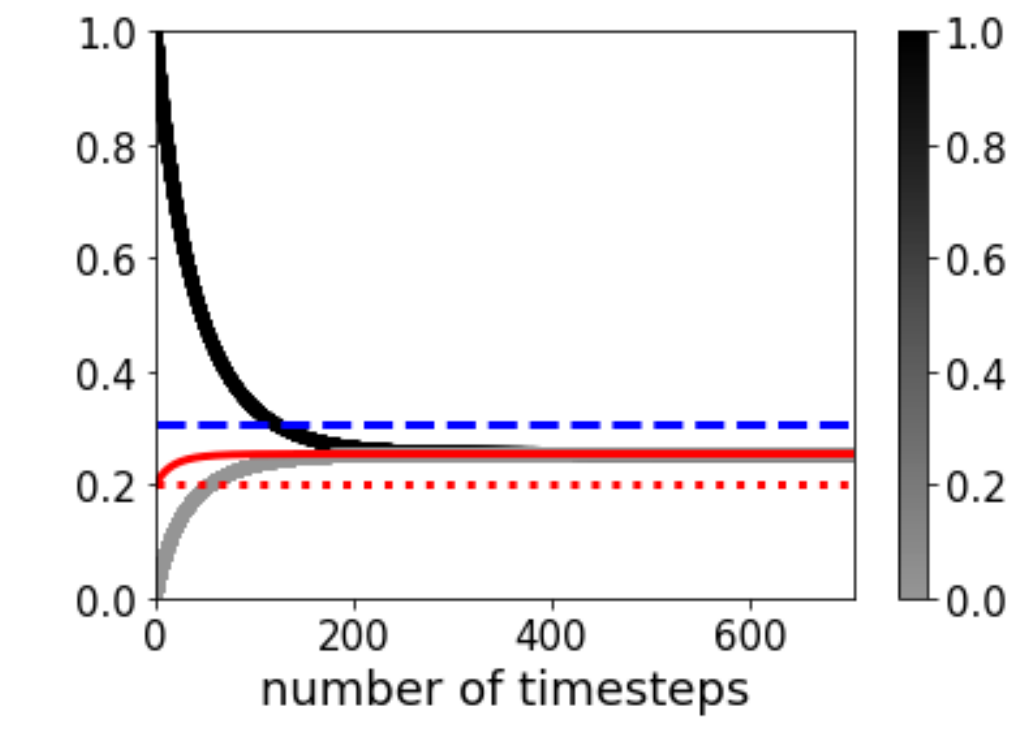}
		\caption{$\lambda=1$ (inhibiting)}
	\end{subfigure}
	\caption{\textbf{Asymmetric initialisation: Conservation only for linear dynamics.}
		An asymmetric initialisation, with $\bar{x}(0)=0.2$, may shift the average node state in 3CM for fully-connected hypergraphs. The scaling function is $s(x)=\exp(\lambda x)$. 
		For $\lambda<0$ (left), the dynamics exhibits a drift towards the majority as similar node states reinforce each other. The opposite effect occurs for $\lambda>0$ (right), as the dynamics exhibits a drift towards balance.
		The average state is conserved for $\lambda=0$ (center), as expected for  linear dynamics. Again, the line colors represent the initial states of the nodes.
		Dotted red lines indicate the initial value of the  average node state. Black (grey) solid lines represent the evolution of the state of nodes whose initial configuration is one (or zero). Dashed blue lines are the final state approximation, $\bar{x}^p$. Reproduced and adapted from \cite{neuhauser_multibody_2020}.
    }
	\label{fig:simulations_meanfield_biased}
\end{figure*}

Considering different initial distributions of the node states, we compared
\begin{enumerate}
    \item a deterministic, symmetric distribution $\mathcal{B}(0.5)$, where $50\%$ of the initial node states have value $0$ and $50\%$ have value $1$ ($\bar{x}(0)=0.5$)
    \item a deterministic, asymmetric distribution $\mathcal{B}(0.2)$, where $80\%$ of the node states have value $0$ and $20\%$ have value $1$ ($\bar{x}(0)=0.2$)
    \item a random initialisation according to a uniform distribution $\mathcal{U}([0,1])$ ($\mathrm{E}(\bar{x}(0))=0.5$)
\end{enumerate}
In the first two cases we do not observe any shift in the average state, as expected. 
We can observe this conservation in \cref{fig:simulations_MeanField_symmetric}, independently of the non-linear dynamics that are applied. 
However, we do see a shift for the asymmetric initialisation, as shown in \Cref{fig:simulations_meanfield_biased}.
The simulations confirm that the average state is conserved for linear dynamics ($\lambda=0$) and multi-way effects only occur for non-linear node interaction functions with $\lambda \neq 0$.
For $\lambda < 0$ we observe a shift of $x(t)$ towards the majority, resulting in an asymptotic average $\lim_{t\rightarrow\infty}\bar{x}(t)$ smaller than the initial value of $\bar{x}=0.2$.
For $\lambda > 0$ we see the opposite phenomenon, with a shift of the average opinion towards $0.5$.

In order to approximate the asymptotic state from the initial configuration of the system, we can use a simple method which estimates the dynamical importance $w_i$ of a node $i$ based on the initial configuration as
\begin{align}
	w_i(t) & =\frac{\text{influence of node } i}{\text{total weight in the system}}                            \nonumber         \\
	       & =\frac{\sum_{j,k=1}^NA^{(3)}_{ijk}s \left( \left| x_i(t)-x_j(t)\right|\right)}{\sum_{i,j,k=1}^N A^{(3)}_{ijk} s \left( \left| x_i(t)-x_j(t)\right|\right)}.
\end{align}
The asymptotic value of $\bar{x}$ is then obtained by one explicit Euler step of the dynamics from the initial configuration  $\bar{x}(0)$ 
\begin{align}
	\label{eqn:predicted_mean}
	\bar{x}^p=\bar{x}(0)+ \sum_{i=1}^Nw_j(0)(x_j(0)-x_i(0)).
\end{align}
The simulations in \Cref{fig:simulations_MeanField_symmetric} and  \Cref{fig:simulations_meanfield_biased}  also display the predicted value \eqref{eqn:predicted_mean}, which  correctly identifies the direction of the shift.

\subsubsection{Analytical examination of MCM I}\label{subsubsec:mcm1_analysis}	
	
Let us now investigate how the shifts in the average node state of the system generalise to the case of non-linear consensus on hypergraphs modeled by the MCM. 
    We will look at the two submodels MCM I and MCM II seperately. 
    Similar to the 3CM, the analytical examination of the MCM II is difficult as the argument of the scaling function $s^{II}_i$, given by the argument function $g_{i \leftarrow j}^{E_\alpha}$, depends on both node $i$ and  node $j$. 
    In contrast, in the case of the MCM I, the argument function $g_{i}^{E_\alpha}$ for the scaling function $s^I_i$ is independent of acting node $j$ and symmetric in all node indices $k \in E_\alpha, k \neq i$. 
    We can thus quantify the shifts of the average opinion in a symmetric system analytically.

	We assume that the scaling function is the same for all nodes and we call it $s^I$. 
    We assume a homogeneous mixing, where the nodes have equal probability of being part of a hyperedge. We consider a hypergraph $\Hy$ with $m_k$ hyperedges of cardinality $k$ for $k=2,3,\hdots,N$. 
    Then, each node participates in $\frac{k m_k}{N}$ hyperedges of cardinality $k$ and that the mean of every hyperedge is the global mean $\bar{x}$.
For an arbitrary node in some hyperedge $i \in E_\alpha$, we thus have
	\begin{equation}
	\dot{x}_i = \sum_{k=2}^{N} \frac{k^2 m_k}{N} s^I\left(\left|\bar{x} - x_i\right|\right) (\bar{x} - x_i)
	\end{equation}
	The mean state evolves as
	\begin{equation}\label{eq:mean evolution}
	\dot{\bar{x}} = \frac{1}{N^2} \left( \sum_{k=2}^N k^2 m_k \right) \left( \sum_{i=1}^N s^I\left(\left|\bar{x} - x_i\right|\right) (\bar{x}-x_i) \right)
	\end{equation}
    and we observe that in a homogeneously mixed system, the mean does not shift if the distribution of $x_i$ about the mean is symmetric. This result is equivalent to the results for 3CM which we examined in Figure \ref{fig:simulations_MeanField_symmetric}.
	
	We can now investigate the effect of an unbalanced initial distribution of the states analytically. Consider a situation where the initial states are binary (either 1 or 0). Suppose at $t=0$, $f_0$ fraction of the nodes have state $0$, and the rest ($f_1 = 1 - f_0$) have state $1$. From Eq.\ref{eq:mean evolution}, we can write
	\begin{align*}
		\dot{\bar{x}} &= \frac{1}{N^2}\left( \sum_{k=2}^N k^2 m_k \right) \left( \sum_{i=1}^N s^I\left(\left|f_1 - x_i\right|\right) (f_1 - x_i) \right)\\
		&= \frac{1}{N}\left( \sum_{k=2}^N k^2 m_k \right) f_0 f_1 (s^I(f_1) - s^I(f_0))
	\end{align*}
    If $s^I$ is monotonically increasing, $f_1 > f_0$ implies that $\dot{\bar{x}} > 0$ and $f_1 < f_0$ that $\dot{\bar{x}} < 0$, i.e. $\bar{x}$ shifts towards the majority. Similarly, $\bar{x}$ shifts towards the minority for monotonically decreasing $s^I$. This is fundamentally different to the case of 3CM, where the same shift appears, but in the opposite direction. We therefore also expect the generalised MCM II to behave in the opposite way to MCM I. In the next section, we thus investigate these contrasting effects, which arise from the different form of non-linear argument.

	\subsubsection{Fundamental differences between MCM I and MCM II}
    To compare the outcomes of the two models, we run numerical simulations on identical topologies (a fully connected hypergraph with $N=10$ nodes) with the same choice of scaling function with different parameters:
	\begin{equation}
	\begin{split}
	s^I_i (x) = s^{II}_i (x) = e^{\lambda x} \;\; \forall i \in V(\Hy)
	\end{split}
	\end{equation}
	As before, the initial node states have binary values ($0$ or $1$),  with $n_0$ nodes of state $0$.
	\begin{figure}
		\centering
		\begin{subfigure}{\linewidth}
			\includegraphics[width=\textwidth]{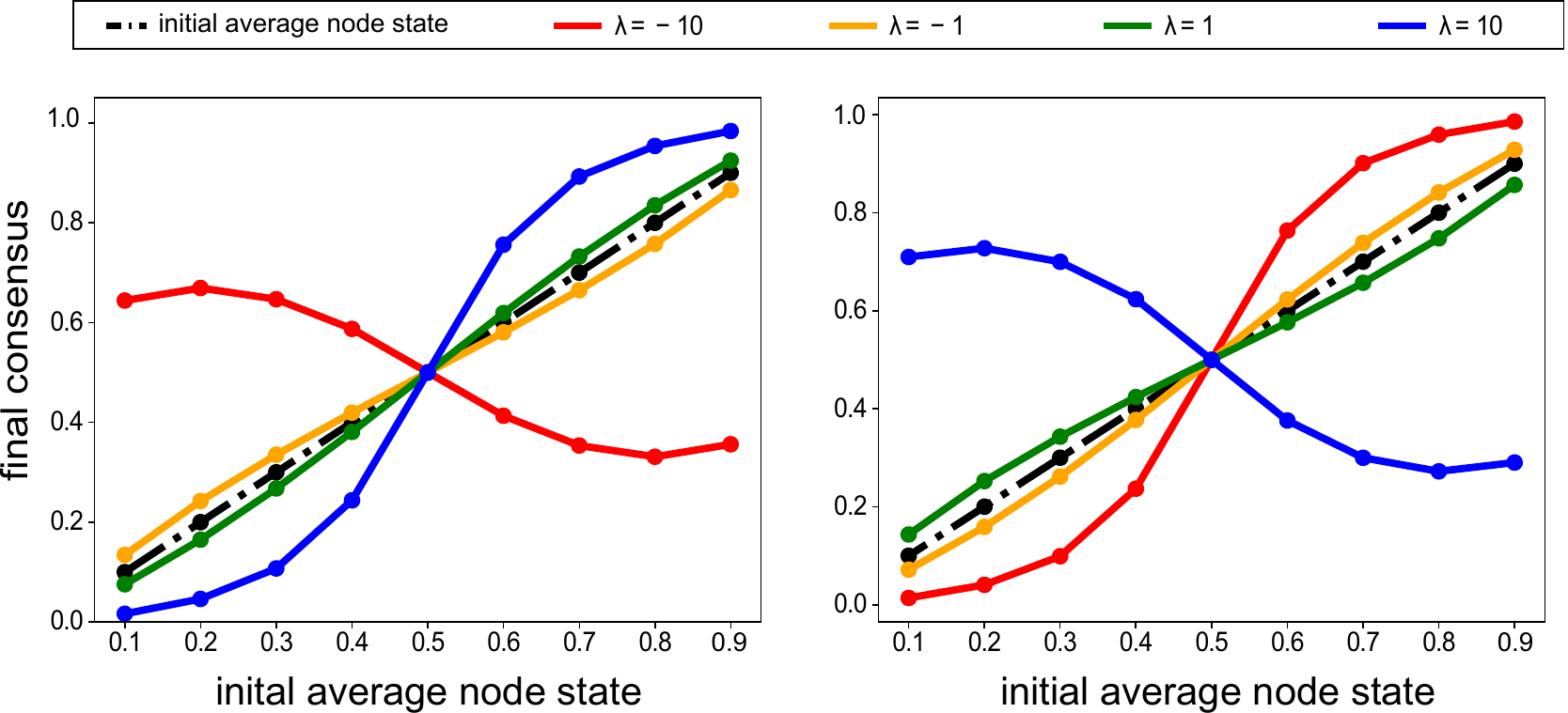}
		\end{subfigure}
		\caption{\textbf{MCM I and MCM II show opposite effects.} Numerical simulations to compare the evolution of MCM I (left) and II (right) on a fully connected hypergraph. Reproduced and adapted from \cite{sahasrabuddhe_modelling_2020}.
        }
		\label{fig:contrast}
	\end{figure}
	
    Numerical results in Fig. \ref{fig:contrast} show that the two submodels MCM I and MCM II evolve in opposite ways. 
    Further, the results for MCM I validate the analytical results in Section \ref{subsubsec:mcm1_analysis}. 
    For a monotonically increasing scaling function $s(x)=e^{\lambda x}$with $\lambda > 0$, we see that the average state of the nodes shifts towards the opinion of the initial majority.
    Similarly, the average state shifts towards the opinion of the initial minority  for a monotonically decreasing scaling function $s^I$ ($\lambda < 0$). 
    While MCM II is a direct generalisation of 3CM and the results of the simulations in Fig. \ref{fig:contrast} (right) align with the numerical results of the previous section, MCM I shows an opposite behaviour to MCM II despite the same scaling function $s^I(x)=s^{II}(x)$.
    These drastic differences underline the huge effect of the argument functions, i.e. $g_i^{E_{\alpha}}$ in the case of MCM I and $g_{i \leftarrow j}^{E_{\alpha}}$ in the cas MCM II, on the long term behavior of the dynamical system.

\subsection{Influence of node specific function parameters}
\label{subsec:node_properties}
\begin{figure}
		\centering
		\begin{subfigure}{0.49\linewidth}
			\includegraphics[width=\textwidth]{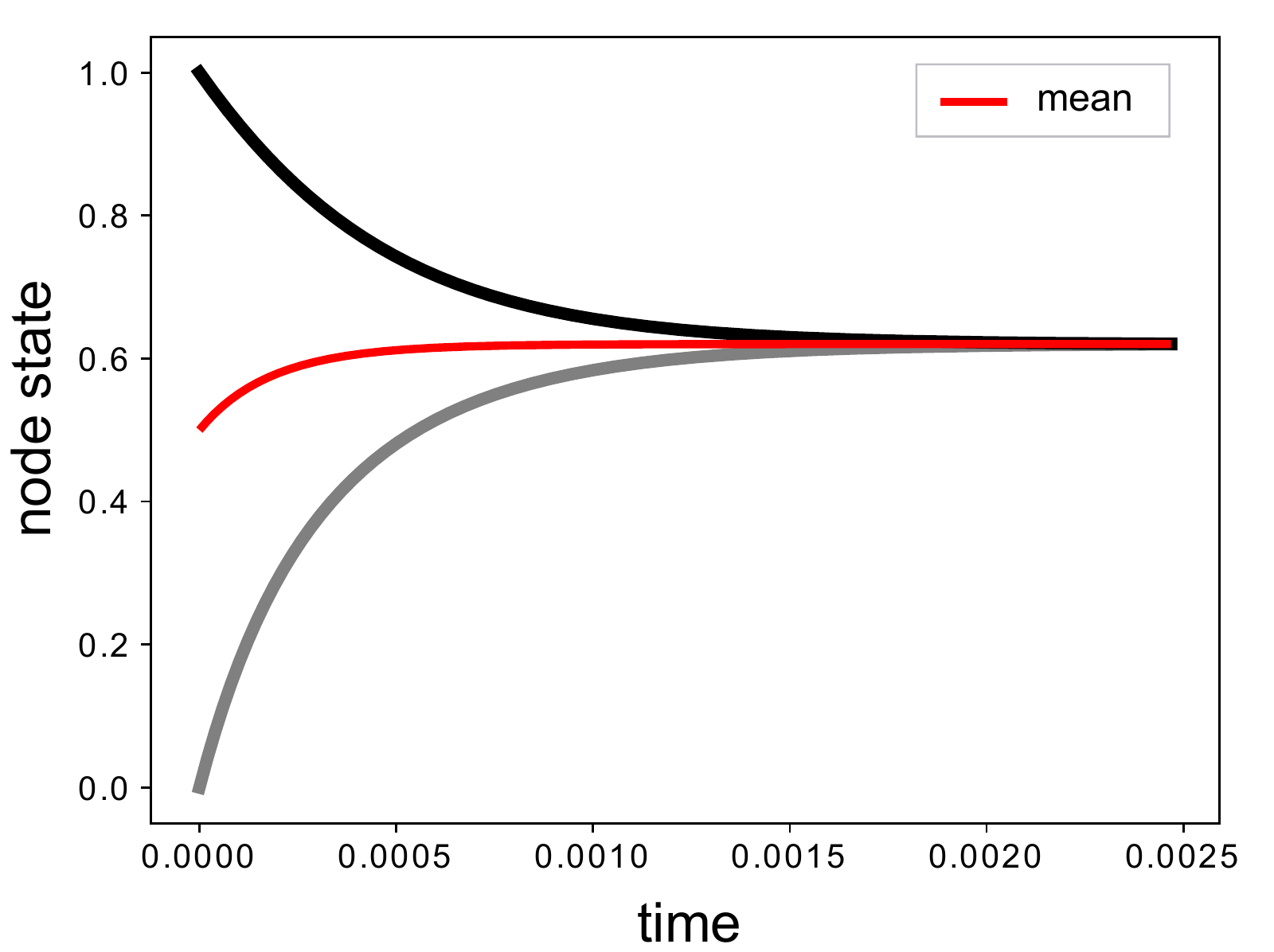}
			\subcaption{Evolution with $\Delta=1$}
		\end{subfigure}
		\begin{subfigure}{0.49\linewidth}
			\includegraphics[width=\textwidth]{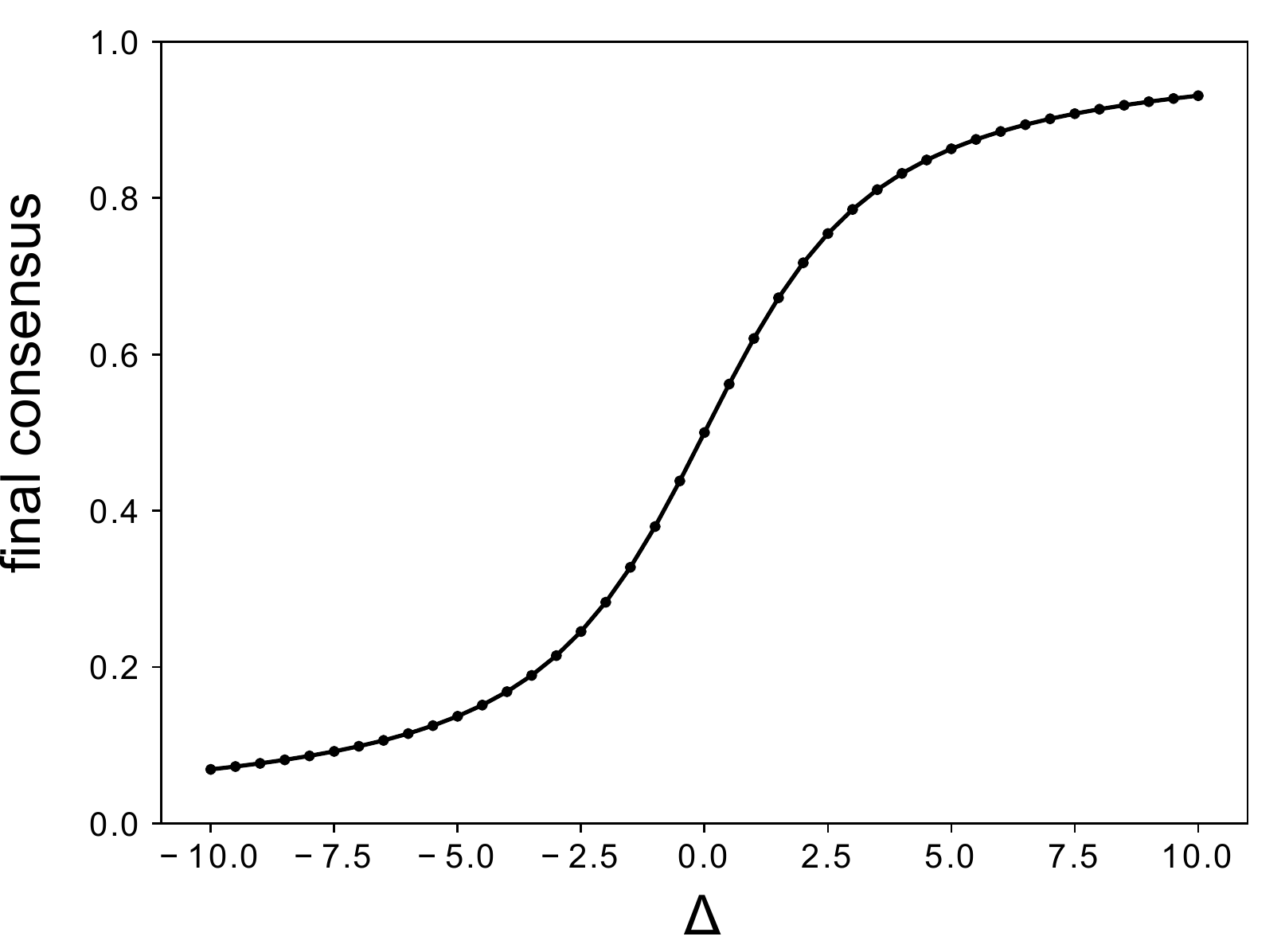}
			\subcaption{Consensus vs $\Delta$}
		\end{subfigure}
		\caption{\textbf{Influence of node specific parameters. }Evolution of MCM I on a fully connected hypergraph of $10$ nodes initialised with $5$ nodes each of opinions $0$ (with $\lambda_i = \Delta$) and $1$ (with $\lambda_i = -\Delta$). Reproduced and adapted from \cite{sahasrabuddhe_modelling_2020}.}
		\label{fig:stubborn_cc}
	\end{figure}

Up until here, we have focused on a dynamics of the form \eqref{eq:mcm1} (MCM I) or \eqref{eq:mcm2} (MCM II/3CM) with  a scaling functions $s(x)$  which was the same for all nodes, e.g. $s_i(x)=s(x)=e^{\lambda x}$ for all $i\in V(\mathcal{G})$. 
However, we can also choose node specific parameters for the scaling functions $s_i(x)$.
These different scaling functions $s_i$ may, e.g., be motivated by sociological aspects such as character traits of different individuals. 
In order to explore the higher-order effects of these individual node traits, consider again for each node $i\in V(\mathcal{G})$ the submodel MCM I \eqref{eq:mcm1} with  exponential function:
	\begin{equation}
	s^I_i(x) = e^{\lambda_i x}.
	\end{equation}
    For $\lambda_i < 0$ ($\lambda_i>0$), the function is monotonically decreasing (increasing). 
    Following the sociological motivation of MCM I in Section \ref{ssec:MCM1}, a decreasing function $s_i$ can model an individual $i$ that is comparably less influenced by groups $E_\alpha$ with an average opinion $\langle{x}_{E_\alpha}\rangle$ that is very different from its own opinion $x_i$, than by groups that have an average opinion similar to its own opinion $x_i$. 
    This can be thought of as individual $i$ resisting change, or some form of 'stubbornness'. 
    On the other hand, an increasing function $s_i$ can be thought of as representing `gullibility', i.e. a rather susceptible individual $i$.
    In Fig. \ref{fig:stubborn_cc}, we present the temporal evolution of the dynamics of the MCM I with $ s^I_i(x) = e^{\lambda_i x}$  on a fully connected hypergraph ($N=10$) with binary, symmetric initialisation.
    The nodes whose states were initialised to $x_i= 1$ (or to initial state $x_i=0$) have a scaling function with parameter $\lambda_i = -\Delta$ (or parameter $\lambda_i=\Delta$), respectively. 
    The numerical results in \ref{fig:stubborn_cc} show that the final consensus of opinion values shifts towards the initial opinion of stubborn individuals.
	
    Note that this is an important aspect of the MCM I, as it enables an individual to heavily influence other members of a group while being resistant to their influence. 
    Hence our models allows certain individuals to be `trendsetters', that can pull entire groups towards their opinion. 
Specifically, stubborn individuals within in a group of people with different opinions will become trendsetters.

  \subsection{Influence of clustered hypergraph topologies}
\label{sec:higher_order_effects:influence_topology}

In the previous experiments we have focused on fully connected hypergraphs, which eliminated the influence of specific hypergraph topologies on  the dynamics. We could thus investigate, how asymmetric distributions of the initial node states cause shifts in the average opinion of the system, which would not be present in a pairwise or a linear higher-order setting. The direction of these shifts were additionally dependant on the form of the scaling function $s_i$. We now want to investigate the additional influence of the hypergraph topology, if it is not symmetric as in the fully connected case. 

If we consider a hypergraph with two clusters $A$ and $B$, we can define two types of hyperedges: If all the nodes of a hyperedge $E_\alpha$ are contained in either cluster $A$ or in cluster $B$, we call $E_{\alpha}$ a \emph{cluster} hyperedge. However, if $E_\alpha$ contains nodes from both clusters, it will be referred to as a \emph{connecting} hyperedge. In the latter case, the connecting hyperedge is called \emph{oriented} towards one of the cluster, if the minority of the hyperedge nodes is part of that cluster. Otherwise, if the nodes are part of cluster $A$ or $B$ in equal numbers, the hyperedge is \emph{unoriented}. 

\begin{figure}
	\centering
	\includegraphics[width=0.8\textwidth]{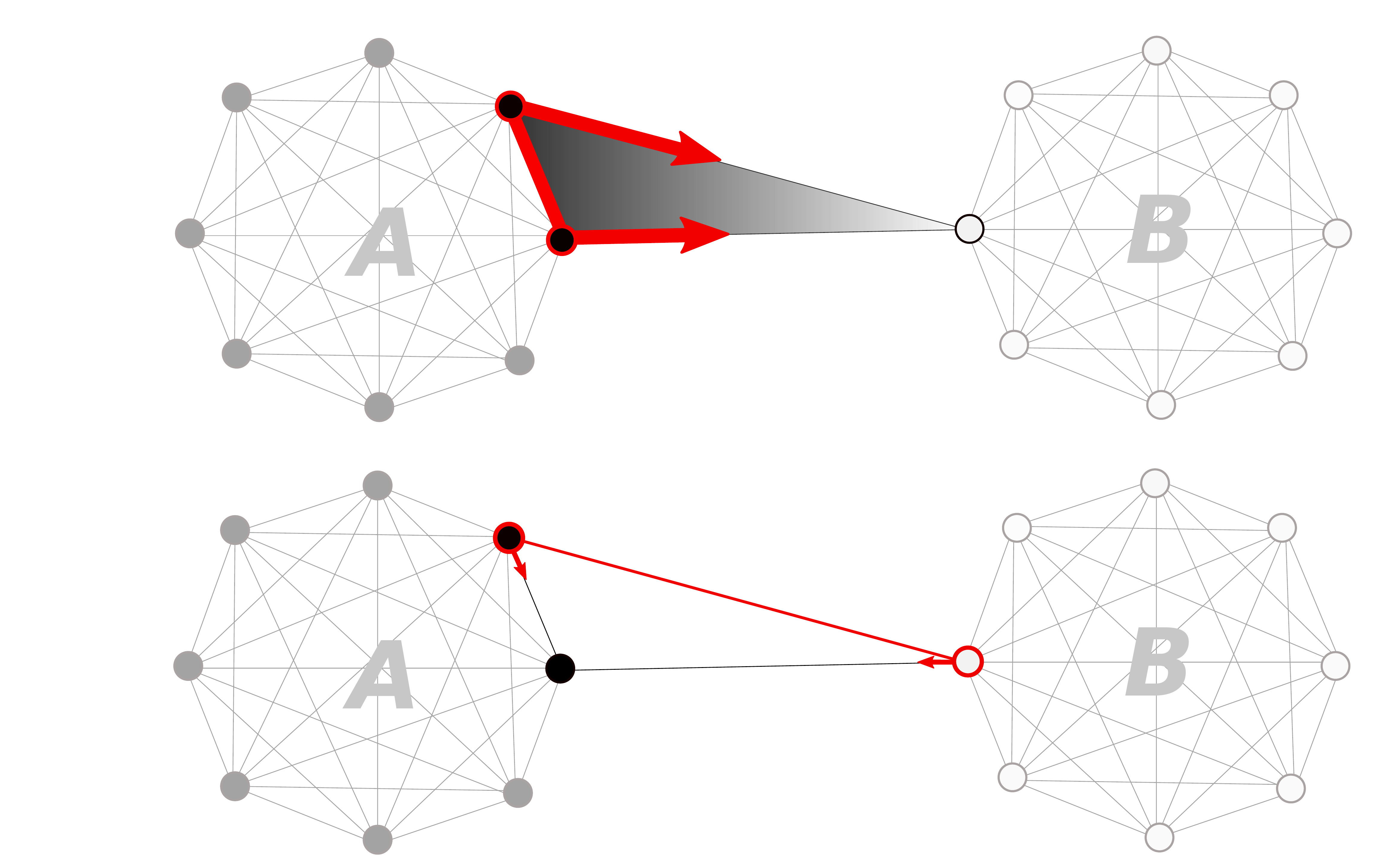}
	\caption{\textbf{Cluster dynamics.}
		If we consider a binary initialisation of the two clusters, here in black and white, and a 3-edge oriented towards cluster $B$ (top), the consensus in cluster $A$ accelerates the rate of change of the neighbour in $B$. 
		In contrast, the node-state difference between the clusters is maximal, which slows down the effect of cluster $B$ on $A$. Reproduced and adapted from \cite{neuhauser_multibody_2020}.
		}
	\label{fig:binary_cluster}
\end{figure}

As an illustrative example we consider a 3-edge hypergraph consisting of two equally sized fully connected clusters.
In addition, we assume that these clusters are connected by a (small) set of 3-edges. This setting is illustrated in Fig. \ref{fig:binary_cluster}.

The dynamical effect of this construction becomes clear if we consider how the initial node states will influence the future dynamics.
We consider binary initial node states, whereas all the nodes in cluster A have the initial state  $x_A(0)=0$ and the nodes in cluster B the initial  state $x_B(0)=1$. 
We consider the 3CM with a positive-definite, decreasing scaling function $s(x)$, such that the influence of nodes with similar states is reinforced within a hyperedge.
Moreover, in this example, we chose to add only a single 3-edge between the cluster $A$ and $B$, that is oriented towards cluster $B$.
Due to the consensus in cluster $A$ and the fact that the connecting hyperedge is oriented towards cluster $B$, the diffusion dynamics is accelerated towards cluster $B$, as the majority of the nodes in the connecting hyperedge have the initial opinion of cluster $A$ and are thus reinforcing each other.
On the contrary, the influence of cluster $B$ is inhibited in the opposite direction, as the node couplings of nodes belonging to different clusters damp the diffusion because of their large state difference.
For this reason, one expects the average initial value in $A$ to dominate that in $B$ and thus to dominate the asymptotic consensus value. 
Note that we thus achieve directed dynamics (or an asymmetric flow) from one cluster to the other.

In order to quantitatively analyse this mechanism, we perform numerical simulations on two fully connected clusters,  each consisting of $10$ nodes, with the binary initialisation  specified above.
We then connect the clusters with $80$ randomly placed $3$-edges, such that a fraction  $p \in [0,1]$ of $3$-edges are oriented towards cluster $A$ and the rest towards cluster $B$.
\subsubsection{Cluster dominance through directedness of cluster connection}
\begin{figure}
	\centering
	\begin{subfigure}{0.45\linewidth}
		\includegraphics[width=\linewidth]{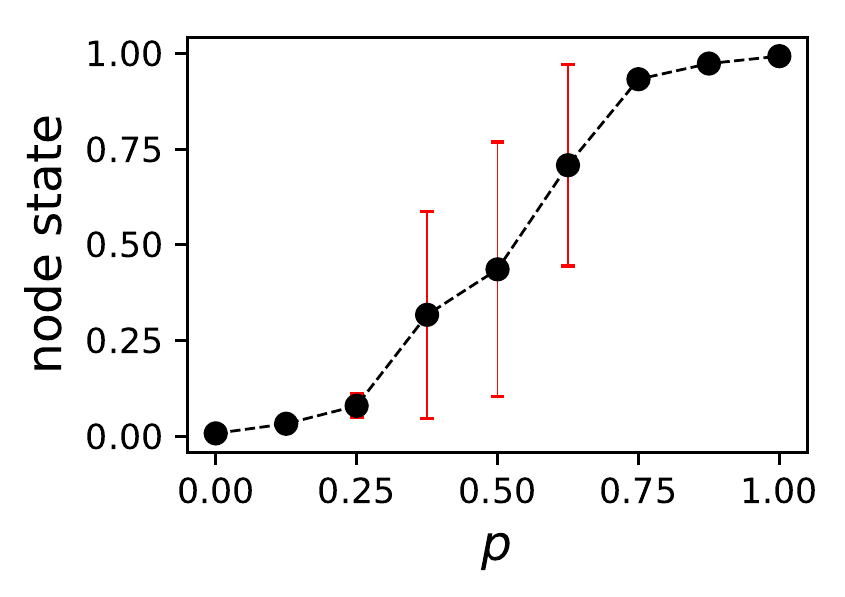}
	\end{subfigure}
	\begin{subfigure}{0.45\linewidth}
		\includegraphics[width=\linewidth]{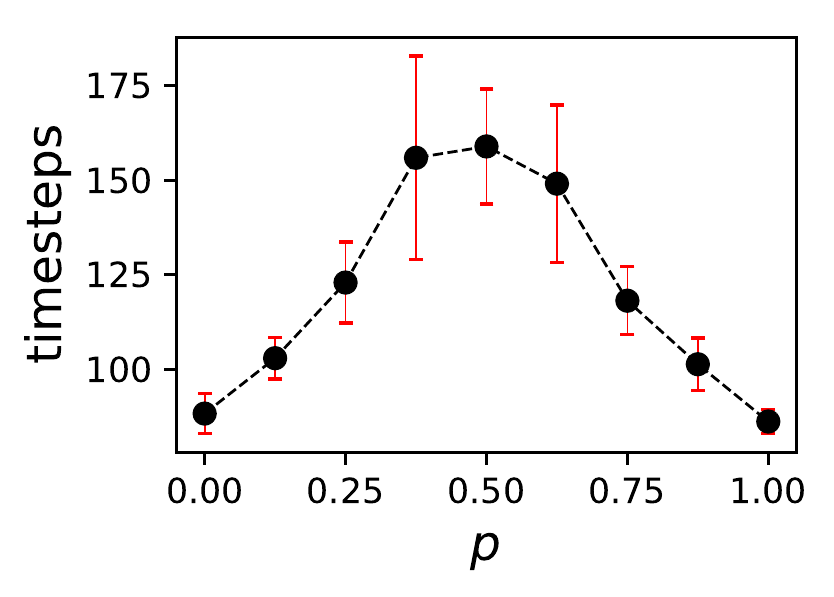}
	\end{subfigure}
	\caption{\textbf{Cluster dominance depends on directedness.}Simulations of 3CM on two inter-connected clusters of 10 nodes, with  the scaling function  $s(x)=\exp(-100x)$ (see main text for a complete description). We compute the final consensus value, averaged over 20 simulations, where the error bars display one standard deviation (left).
		As the fraction of 3-edges directed from cluster A to cluster B increases, so does the consensus value towards the initial state in cluster A.
		The rate of convergence is significantly faster when the initial configuration is very asymmetric, that is extreme values of $p$ (right). Reproduced and adapted from \cite{neuhauser_multibody_2020}.
		}
	\label{fig:p_exp}
\end{figure}

We first  examine the influence of the orientation parameter $p$.
For that purpose, we take the scaling function $s(x)=\exp(\lambda x)$ with $\lambda =-100$, so that pairs of similar nodes exert a strong influence on other nodes.
We show the results of our model simulations averaged over $20$ random instances in \Cref{fig:p_exp}.
In \Cref{fig:p_exp} (left), we observe a shift in the final consensus value towards the initial value in cluster $A$ (or cluster $B$, respectively). 
The direction of this shift depends on the orientation of the connecting 3-edges, quantified by $p$. For $p=0$, all the connecting 3-edges are oriented towards cluster $B$ and the initial opinion of cluster $A$ thus dominates the dynamics. On the contrary, for  $p=1$ all connecting traingles are oriented towards cluster $A$, therefore we observe a maximal influence of the initial value of cluster $B$.

The asymmetry of the dynamics which results from the orientation of the connecting hyperedges also influences the rate of convergence towards consensus, as shown in \Cref{fig:p_exp} (right).
More asymmetric configurations lead to a faster rate of convergence.
The simulations also reveal higher fluctuations in the asymptotic state for values close to $p=0.5$.
This result indicates that the process is  sensitive to even small deviations from balance in the initial topology, which can lead to large differences in the consensus value.

\begin{figure}
	\centering
	\begin{subfigure}{0.6\linewidth}
		\includegraphics[width=\linewidth]{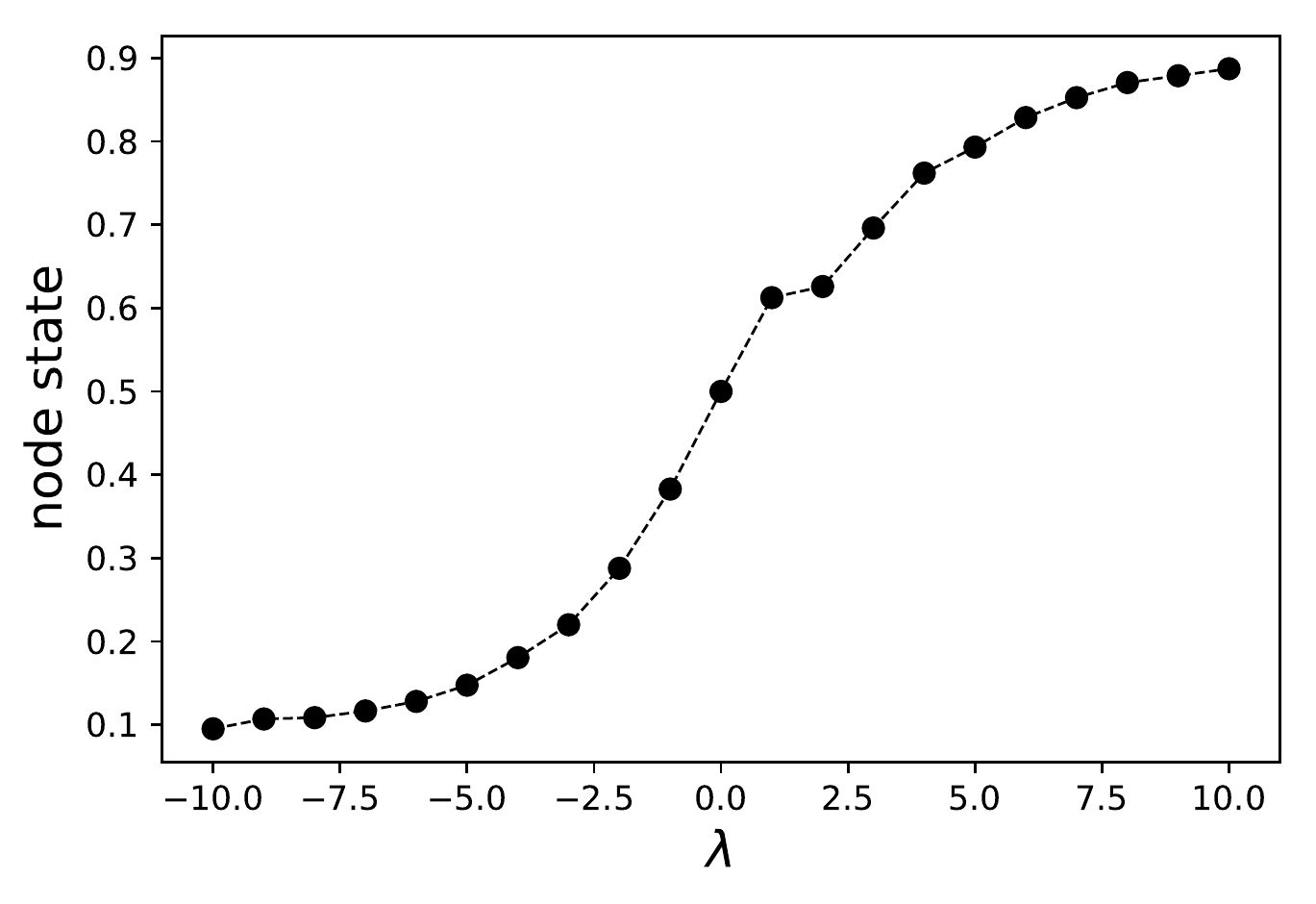}
	\end{subfigure}
	\caption{\textbf{Cluster dominance for reinforcing vs. inhibiting dynamics.}
	The final consensus value of the two cluster system (with $p=0$), dependent on the parameter $\lambda$.
	As the connecting 3-edges are all oriented towards cluster B, it depends on $\lambda$ if the nodes in cluster A are reinforcing each others' influence which leads to directed dynamics towards cluster B ($\lambda<0$) or inhibit each other ($\lambda>0$) which leads to the contrary effect.
	Therefore, the consensus value shifts towards the mean of cluster B with growing $\lambda$.
	For $\lambda=0$ we have linear dynamics and the initial average $0.5$ is conserved. Reproduced and adapted from \cite{neuhauser_multibody_2020}.
	}
	\label{lambda_exp}
\end{figure}

Note that the effect of the orientation reverses if we consider an increasing scaling function such that dissimilar node states reinforce each other. In \Cref{lambda_exp}, we examine this effect by changing $\lambda$ for the scaling function $s(x)=\exp(\lambda x)$. We observe a transition from the initial value in cluster $A$ to that in cluster $B$, as expected. 

\subsubsection{Minority influence}
\begin{figure}
		\centering
        \begin{subfigure}{0.49\linewidth}
		\includegraphics[width=\linewidth]{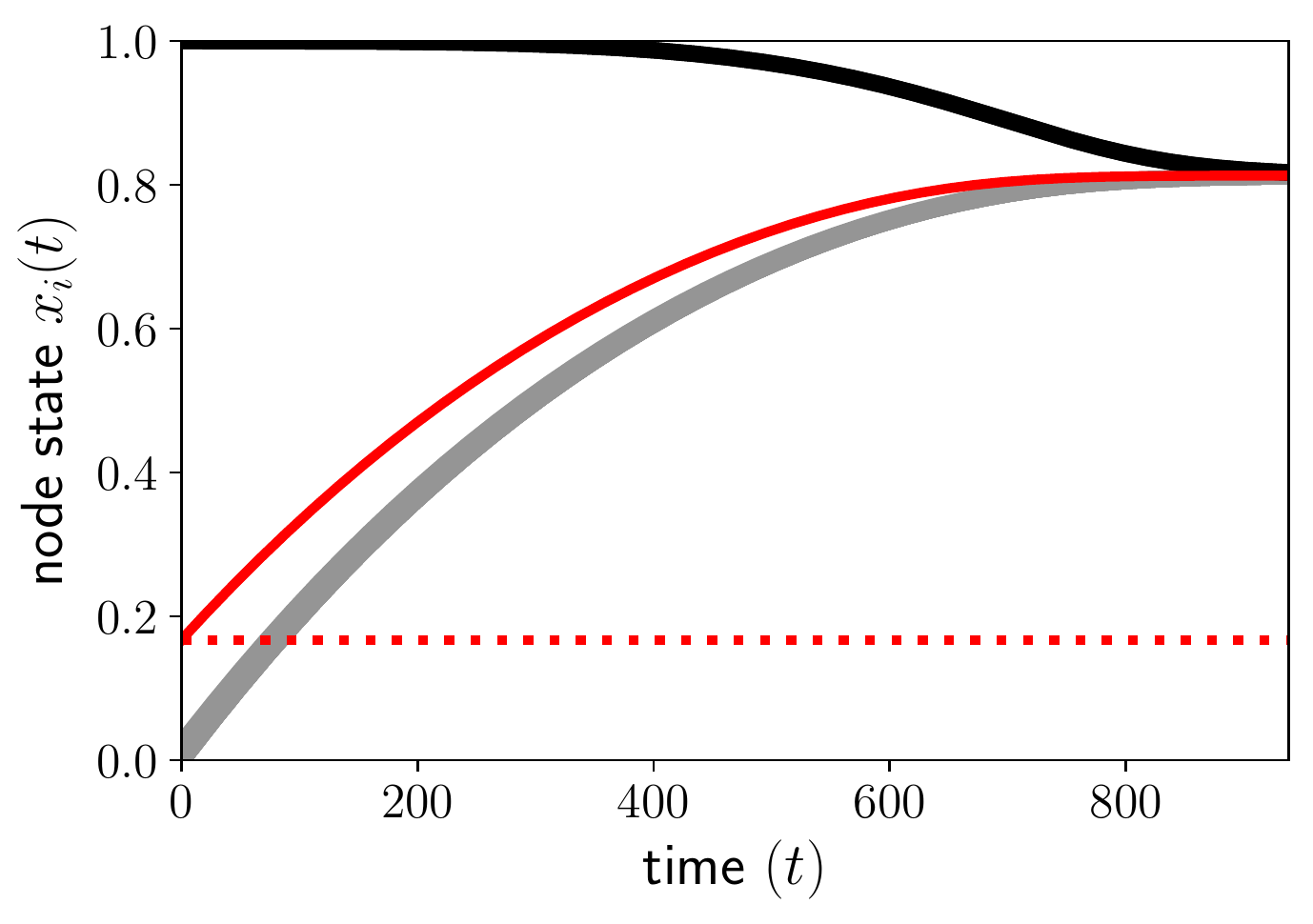}
        \end{subfigure}
        \begin{subfigure}{0.49\linewidth}
		\includegraphics[width=\linewidth]{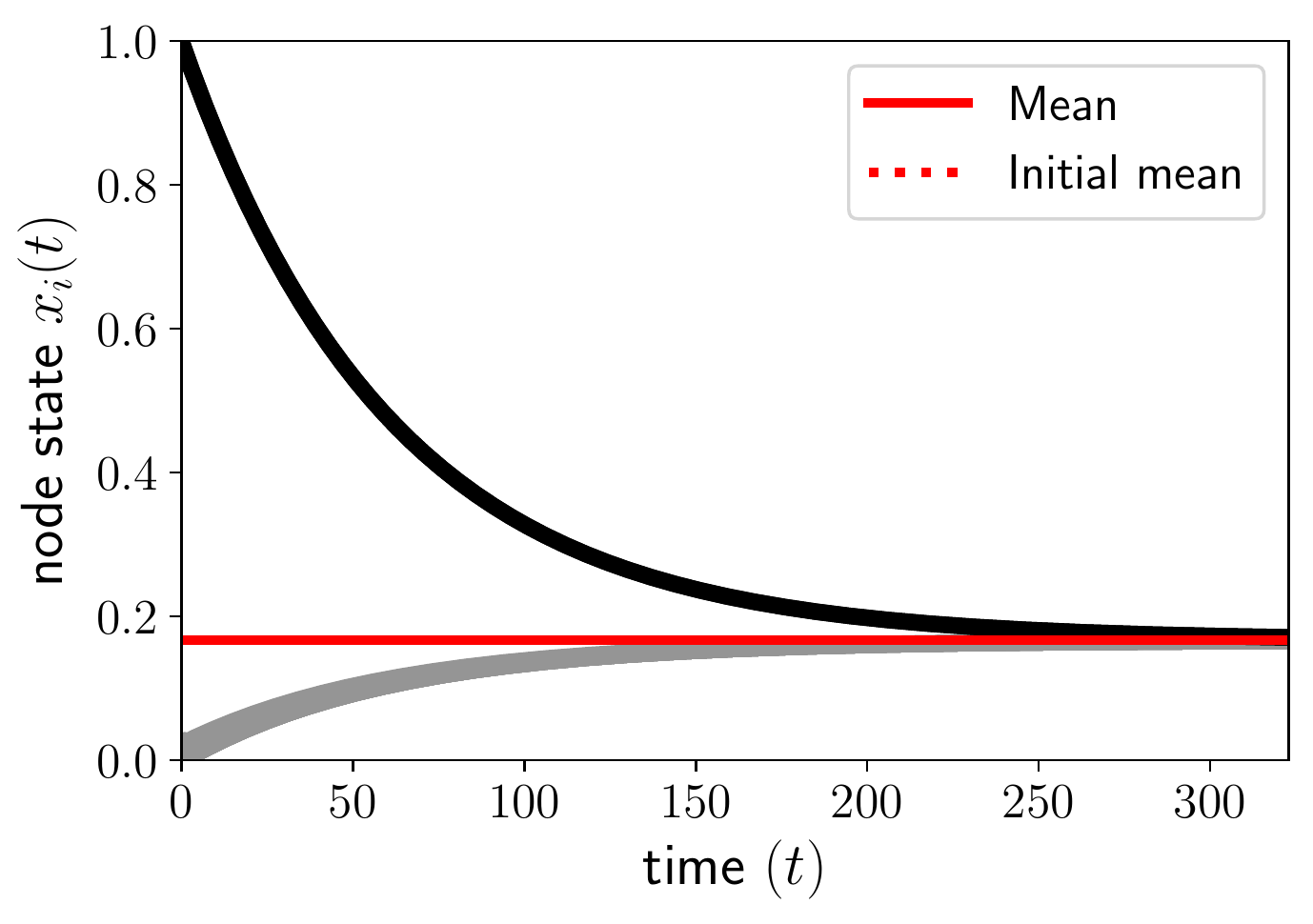}
        \end{subfigure}
		\caption{
			\textbf{Global minority influence through reinforcing opinion.}
			We display simulations for a scaling function $s(x)=\exp(\lambda x)$ for $\lambda=-10$ and two clusters of different sizes, which are connected with a single 3-edge oriented towards cluster A. 
            Cluster A comprises the majority of nodes (10 nodes) whereas cluster B consists only of 5 nodes. While intuition may suggest a final consensus that is leaning towards the initial opinion $0$ of the majority cluster A, we observe the opposite behavior due to opinion reinforcing effect of the nonlinear coupling, which leads to an (effectively) directed dynamics between B and A (left). In contrast, if the dynamics are linear (right), the initial average opinion is conserved and therefore the majority opinion dominates the final consensus value. Reproduced and adapted from \cite{neuhauser_opinion_2020}.
		}
		\label{fig:minorities}
	\end{figure}

Instead of considering equally sized groups as in the last section, we can also consider settings in which one cluster forms a ``minority'' and is comparably smaller than the other cluster (the majority) \cite{neuhauser_opinion_2020}. 
As shown in Figure~\ref{fig:minorities}, even in this case the opinion of the global minority may have a stronger influence on the final consensus value than the majority cluster, depending on the relative number of 3-edges oriented towards the majority. 
In the context of opinion dynamics, this type of behavior is akin to a ``minority influence'', where small groups can dominate the formation of an opinion.
Note that this happens not because of the size of the minority group, but due to the internal cohesion of opinions within the minority and because the minority nodes form the local majority in the connecting subgroups. 
Accordingly, if the minority does not agree on the same opinion or the connecting subgroups are not oriented towards the majority, the minority influence is diminished.
Likewise, if we remove the nonlinear effect of opinion reinforcement via the scaling function $s$ and consider simply a linear coupling, then the initial opinion of the majority will have the strongest effect on the final consensus state.

\subsubsection{Heaviside function: Bounded confidence models on hypergraphs}
Up until now, the scaling function $s$ has always been an exponential function, $s(x)=\exp(\lambda x)$, to demonstrate general properties of the model. 
However, we can choose other functions as scaling functions. 
One interesting option is the Heaviside function, given by
\begin{align}
	s(|x_j-x_k|)=H(|x_j-x_k|-\phi)=\begin{cases}0 & \mbox{if } |x_j-x_k|<\phi \\
		1 & \mbox{otherwise},
	\end{cases}
\end{align}
which switches between a zero interaction and linear diffusion when the difference of the neighbouring nodes becomes smaller than a threshold $\phi \in (0,1)$.
This property is reminiscent of the bounded confidence model \cite{blondel_krauses_2009}.
Note that the Heaviside function is not positive-definite, so that nonlinear consensus dynamics (MCM I and II) with this scaling function do not necessarily converge to a consensus state asymptotically, in which all nodes have the same value.

\begin{figure}
	\centering
	\includegraphics[width=0.7\linewidth]{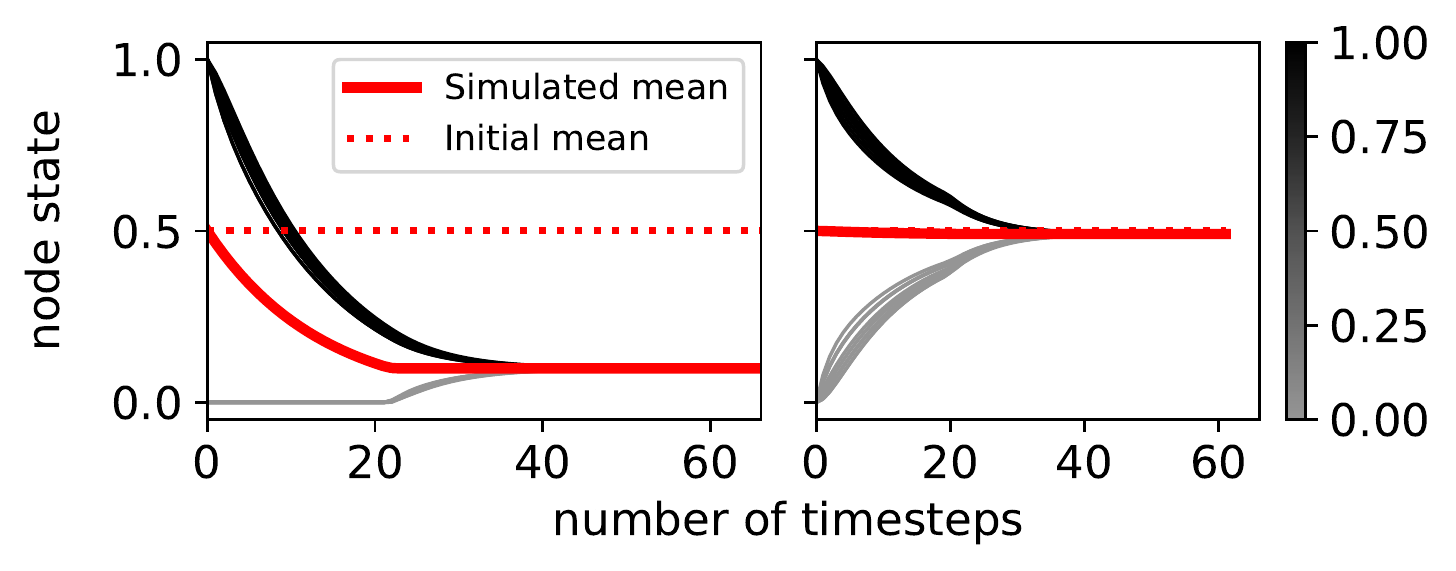}
	\caption{\textbf{Bounded confidence on hypergraphs. }Time evolution of the node states for a Heaviside function with $\phi=0.2$.
		For $p=0$ (left), only diffusion from cluster $A$ towards $B$ is enabled, until the threshold of the Heaviside threshold $\phi=0.2$ is reached.
		The dynamics then become linear and the average state becomes conserved.
		For $p=0.5$ (right), the dynamics are initially symmetric, as the orientation of the connecting 3-edges are balanced, and the dynamics are simultaneously switched on. Reproduced and adapted from \cite{neuhauser_multibody_2020}.}
	\label{fig:heaviside_exp}
\end{figure}

In \Cref{fig:heaviside_exp}, we show the simulation results for $\phi=0.2$ and a cluster scenario with $p=0$, i.e., the 3-edges are all oriented towards cluster $B$.
As the difference of the node states of the two clusters is initially larger than $\phi$, only the nodes of cluster $A$ in the connecting 3-edges are close enough such that linear diffusion takes place towards cluster $B$.  
Therefore, only nodes of cluster $B$ change their value initially as shown in \Cref{fig:heaviside_exp} (left). 
As soon as the difference of the node states of the two clusters is smaller than $\phi$, the opinion dynamics is switched on for all node couplings and the dynamics becomes linear. Therefore, the asymmetry of the dynamics disappears. 
This mimics a situation of two polarised clusters where one side makes unilateral concessions until the other side starts to participate in the consensus formation again. 
For $p=0.5$ the dynamics are symmetric as the orientation of the 3-edges is balanced.

\begin{figure}
	\centering
	\begin{subfigure}{0.8\linewidth}
		\includegraphics[width=\linewidth]{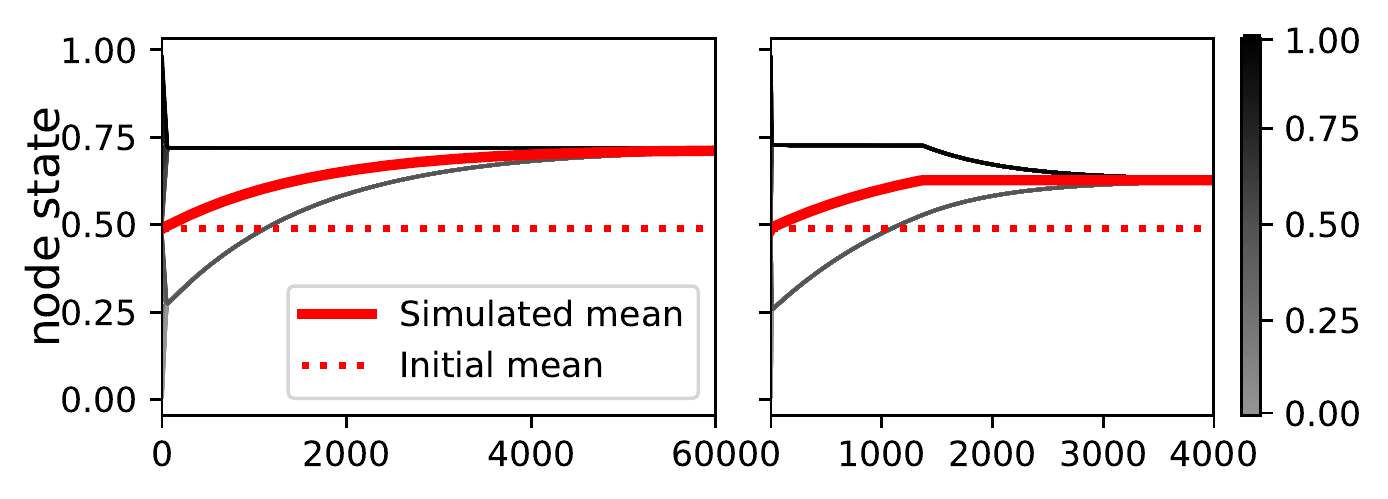}
	\end{subfigure}
	\begin{subfigure}{0.8\linewidth}
		\includegraphics[width=\linewidth]{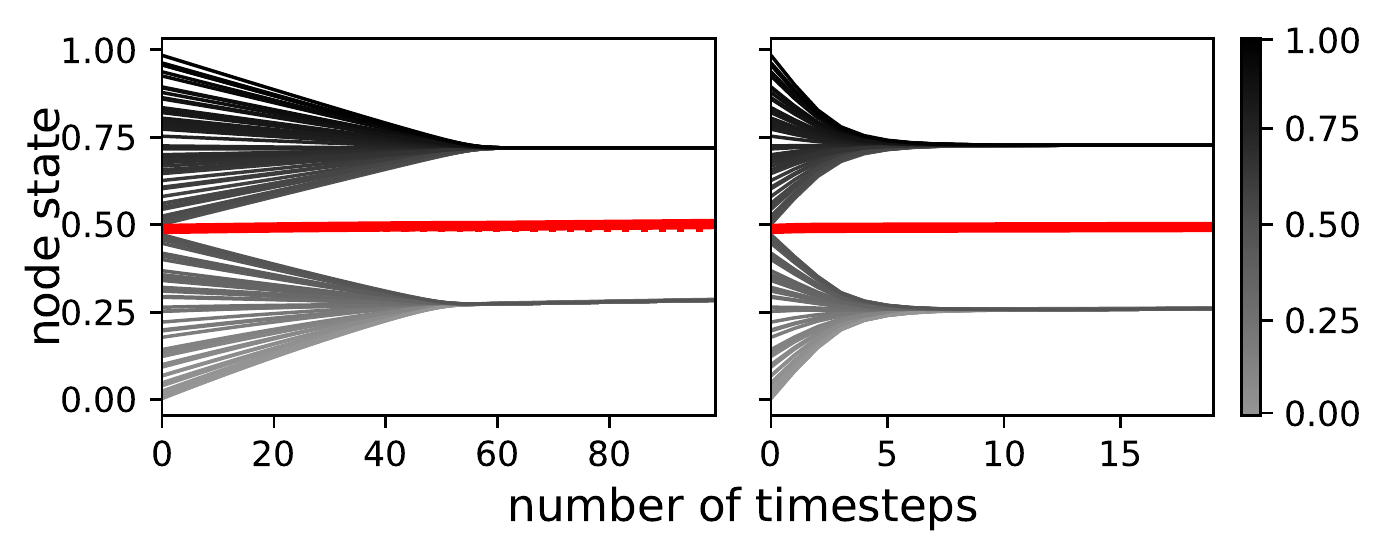}
	\end{subfigure}
	\caption{\textbf{Time-scale separation. }
		Dynamics of two clusters $A$ and $B$ connected with $p=1$, e.g. with 3-edges oriented towards nodes in $A$, and initialised with uniform distributions over separate intervals $I_A, I_B$ with $I_A \cap I_B = \emptyset$. The left figures correspond to the exponential scaling function $s(x)=\exp(\lambda x)$ for $\lambda=-100$, and the right to a Heaviside function with threshold  $\phi=0.2$.
		We observe a timescale separation with a fast, symmetric dynamics inside the clusters, followed by a slow, asymmetric dynamics between the clusters. The fast dynamics is shown in the bottom figures, with qualitatively similar results for both scaling functions. The top figures show a shift towards cluster $B$ for the slow dynamics. For the Heaviside function, the process becomes linear when the node states in the two clusters are less separated than the Heaviside-threshold. Reproduced and adapted from \cite{neuhauser_multibody_2020}.
	}
	\label{fig:timescale_sep}
\end{figure}

\subsubsection{Time-scale separation}
Finally, we investigate our multi-way interaction dynamics concerning the interplay between the topology in a clustered hypergraph and initial conditions that is not bimodal.
    In particular, we are interested in examining different time scales in the dynamics induced by the clustered topology with two groups.
    The different time-scales are here associated to a fast convergence of states inside the clusters, followed by a slower convergence towards global consensus.

    Specifically, we reconsider the clustered 3-edge hypergraph with $p=1$ and with a dynamics governed by the 3CM.
    This time, however, the nodes in each cluster have different states initially.
    For our experiments, we initialise nodes in different clusters uniformly at random over disjoint intervals, such that nodes of cluster $A$ have random initial states in the interval $I_A=[0,0.5]$ and nodes in cluster $B$ have random initial states in $I_B=[0.5,1]$ (see \Cref{fig:timescale_sep}).
	The initial cluster averages of the node states are thus separated.

    Two effects lead now to a fast multi-way consensus inside each cluster.
	First, each of the clusters are internally fully-connected.
    Second, the inter-cluster-dynamics will have a weaker effect, since the difference in the distribution of the initial conditions implies that $s\left(\left|x_i - x_j\right|\right)$ will be small if nodes $i$ and $j$ are in different clusters. 
    As a result, we first observe a fast dynamics within the clusters in which nodes approach the cluster-average state (\Cref{fig:timescale_sep}, bottom) and then a slower dynamics between the two clusters (\Cref{fig:timescale_sep}, top).
    
    The observed effect of the dynamics additionally depends on the scaling function. 
    For an scaling function $s(x)=\exp(\lambda x)$, with $\lambda=-100$, we observe an asymmetric shift of the average node state as shown in \Cref{fig:timescale_sep} (left). 
    If we consider the Heaviside function as a scaling function $s(x)$ instead, the dynamics show a similar asymmetry as in the exponential case, until the two cluster means are less separated than the Heaviside threshold $\phi=0.2$. 
    As shown in \Cref{fig:timescale_sep} (right), the dynamics then become linear and symmetric.

\label{sec:Higher-order-effects}
\section{Conclusion}

In this chapter, we have emphasised the importance of non-linearity for a dynamical process to exhibit higher-order effects. 
Specifically, our results show that  it is important to distinguish between the model of the multi-body structure of a system (here: a hypergraph) and the model of its multi-way dynamics (here: non-linear multi-way consensus). 
The interplay of both aspects is important for genuine higher-order effects to emerge. 
This is particularly apparent for linear consensus models, whose dynamics can always be reduced to a pairwise dynamical system even when defined on a hypergraph. 
In other words, it is a necessary (but not sufficient) condition that the node interaction function is non-linear for genuine, non-reducible multi-way dynamical phenomena to emerge. 
In that case, adjacency matrices are not adequate to encode that sub-groups of nodes interact together, and higher-order objects like hypergraphs are required. 

We have then analysed possible higher-order dynamical effects by looking at specific non-linear interaction functions, which are inspired by models in opinion dynamics. 
We introduced a general Multi-way consensus model (MCM) in which the adoption of an opinion by a node within a group is scaled non-linearly by the similarity of the group members, either including or excluding the affected node. 
This leads to submodels, the MCM I and the MCM II, which have different mathematical properties due to the dependencies of the argument of the non-linear scaling. 
In sociological terms, the two submodels can represent consensus dynamics that are either driven by homophily (MCM I) or by conformity or peer-pressure (MCM II). 

The resulting dynamics lead to shifts of the average opinion state in the system, which would not be present in the case of pairwise or linear multi-way interactions.

In a fully connected system, we find that the shift in the final consensus value only depends on the interplay between i) the distribution of the initial states of the nodes (no shifts if $\bar{x}(0)=0.5$ vs. shifts for $\bar{x}(0)\neq 0.5$) and ii) the form of the non-linearity of the dynamics, i.e., the scaling function $s(g(x))$ (reinforcing (inhibiting) dynamics for monotonically decreasing (increasing) $s$ in the case of 3CM, opposite effects of MCM I and MCM II due to the different form of their argument function $g$).
If we additionally consider a scaling function $s_i$ with node specific parameters, which classify certain nodes to be more stubborn than others (in the case of MCM I) we can observe shifts even in the case of $\bar{x}(0)= 0.5$.

If we go beyond fully connected systems and thus additionally consider the influence of the hypergraph topology, we observe that the influence of certain node subsets can dominate the final consensus value in clustered hypergraphs. This depends on the orientation of the hyperedges connecting the cluster. 
In the case of bounded-confidence dynamics, this mechanisms can even lead to a situation in an opinion dynamics where only one subgroup makes concessions initially. Moreover, in the case of initial state distributions which are not bimodal, a combination of symmetric and asymmetric dynamics is possible: we observe a timescale separation with a fast, symmetric dynamics inside the clusters, followed by a slow, asymmetric dynamics between the clusters.

Generally, we conclude that non-linearity is needed for higher-order dynamical effects to appear on hypergraphs. 
The effects that appear depend on a complex interplay between the type of the non-linear dynamics, the topology of the hypergraph and the initial node states. 
We have explored this interplay for a family of models for consensus dynamics, and the rich phenomenology that we observed motivates the study of these questions for other families of dynamical models in the future.

\label{sec:conclusion}
\subsection*{Acknowledgements}
We acknowledge partial funding from the Ministry of Culture and Science (MKW) of the German State of North Rhine-Westphalia (NRW Rückkehrprogramm).

\bibliographystyle{plain}
\bibliography{references}
\end{document}